\PassOptionsToPackage{table,prologue}{xcolor}

\documentclass[acmsmall]{acmart}
\AtBeginDocument{%
  }

\acmJournal{TOSEM}

\usepackage[utf8]{inputenc}
\usepackage{array}
\usepackage{wrapfig}
\usepackage{multirow}
\usepackage{tabularx}
\usepackage{caption}
\usepackage{subcaption}
\usepackage{booktabs}
\usepackage{setspace}

\usepackage{amssymb}
\usepackage{enumitem}
\usepackage{multirow}
\usepackage{longtable}
\usepackage{changepage}
\usepackage{hyperref}
\hypersetup{
    colorlinks=true,
    linkcolor=black,
    filecolor=black,      
    urlcolor=black,
    citecolor=black,
}
\usepackage{graphicx}
\usepackage{longtable}
\usepackage{adjustbox}
\usepackage{comment}

\newcommand{\toolname}[0]{}
\renewcommand{\toolname}[0]{{\textit{AdaptForge}}}

\begin{document}

\title{Developer Perceptions on Utilising Low-Code Approaches to Build Accessible and Adaptive Applications for Seniors}

\author{Shavindra Wickramathilaka}
\email{shavindra.wickramathilaka@monash.edu}
\affiliation{%
  \institution{Faculty of Information Technology, Monash University}
  \city{Melbourne}
  \state{Victoria}
  \country{Australia}
}

\author{John Grundy}
\affiliation{%
  \institution{Faculty of Information Technology, Monash University}
  \city{Melbourne}
  \country{Australia}}
\email{john.grundy@monash.edu}

\author{Kashumi Madampe}
\affiliation{%
  \institution{Faculty of Information Technology, Monash University}
  \city{Melbourne}
  \country{Australia}}
 \email{kashumi.madampe@monash.edu}

\author{Omar Haggag}
\affiliation{%
 \  \institution{Faculty of Information Technology, Monash University}
  \city{Melbourne}
  \country{Australia}}
\email{omar.haggag@monash.edu}


\begin{abstract}

The global ageing population presents a growing societal challenge, creating an urgent need for inclusive technologies that promote autonomy among older adults. Software practitioners can address this by delivering digital services that enhance seniors' independence and reduce reliance on routine support from family members and healthcare infrastructure. However, traditional development practices, constrained by time and resources, often result in applications with major accessibility and personalisation barriers. Increasing pressure from regulatory requirements, such as the European Accessibility Act (EAA), and the personal empathy many developers feel toward supporting their older loved ones and their own future selves have created a demand for tools that support the development of accessible and adaptive software. To address this demand, this paper presents an interview-based empirical study with 18 software practitioners, evaluating \toolname: a low-code model-driven engineering (MDE) tool that enables the efficient creation of accessible and adaptive applications for senior users by mitigating development constraints through automated code generation. Based on these insights, we identify developer expectations for adopting such tools as industry-standard solutions and provide empirically grounded recommendations for designing low-code tools that support accessible and adaptive software development.


\end{abstract}




\maketitle

\section{Introduction}\label{introduction}

The ageing population represents one of the most significant challenges confronting our society~\cite{who2024}. However, focusing exclusively on its perceived negative implications, such as diminished global productivity or portraying seniors as a societal burden, is both ageist and discriminatory~\cite{who2024}. Instead, we should highlight the positive aspects of longevity, including the ability of older adults to spend extended time with their families and continue making meaningful contributions to society. Nonetheless, recognising these benefits does not negate the necessity of addressing the genuine challenges associated with population ageing, which require thoughtful and inclusive societal responses.

In this context, a promising strategy to alleviate the pressure on support infrastructures for seniors is the strategic integration of technology to supplement and enhance care and autonomy for older individuals~\cite{cheek2005}. For example, many seniors express a preference to live independently in familiar settings, such as a home that they lived in for a long time, rather than transitioning to managed healthcare environments~\cite{cheek2005}. In such a context, the ability to perform tasks such as banking, shopping, and socialising through digital devices and services, from the comfort of their homes or any preferred location, remarkably enhances their sense of autonomy. Additionally, it reduces their dependency on family members and the healthcare system for routine support.

Therefore, it is unfortunate that software practitioners, in attempting to deliver digital services to seniors, have often contributed to the creation of accessibility barriers. Instead of enhancing convenience, these applications frequently result in significant frustration for the elderly community~\cite{wickramathilaka2025guidelines, bossini2014, connor2017, Paez2019}. It is important to recognise, however, that software practitioners are also individuals who often empathise with the challenges faced by seniors \cite{wickramathilaka2025guidelines, vu2022better}. Many have witnessed their own older relatives struggle with software accessibility and may have offered support to overcome such barriers. Yet, this personal empathy does not always translate into professional practice due to constraints such as tight project deadlines and limited development resources \cite{wickramathilaka2025guidelines}.

This is where low-code tool developers can offer meaningful solutions. Model-Driven Engineering (MDE) is an emerging paradigm in which abstract modelling artefacts, expressed through general-purpose languages (e.g., UML) or domain-specific languages (DSLs), are automatically transformed into source code through code generation~\cite{brambilla2017}. As a low-code approach, MDE allows software practitioners to specify the accessibility and personalisation requirements of specific user groups, such as older adults, using a custom domain-specific language suite. Rather than manually implementing accessibility features or creating customised application instances for each user or group, practitioners can delegate these tasks to the MDE code generation process. This automation enables the generation of accessible and personalised application instances tailored to the users' needs~\cite{wickramathilaka2025technical, yigitbas2020, bendaly2018, minon2015}. Moreover, this process requires minimal developer intervention, allowing practitioners to allocate their time to other value-added tasks.


To support this automation need, we developed a low-code tool named \toolname, specifically aimed at addressing the diverse and personalised needs of seniors (older adults over the age of 60)~\cite{wickramathilaka2025technical}. This tool was positively received in a user study that gathered evaluative insights from both software developers and senior end users~\cite{wickramathilaka2025technical, wickramathilaka2025guidelines}. The developer perspective was central to this study, as \toolname~was explicitly designed with software practitioners as the primary users. Accordingly, we conducted an interview-based study involving 18 professional software developers from 5 different continents, with diverse levels of experience as software practitioners\footnote{Approved under Monash University Human Research Ethics Committee (MUHREC), Project ID: 42470}. Each interview lasted approximately one hour. The study yielded several important insights on how to enhance \toolname~to better support real-world software development practices. These included improving the ease of use of the modelling interface, refining the code generation process, and ensuring that the MDE automation supports key software life cycle activities such as code readability, maintainability, application deployment, and quality assurance, rather than compromising them.


In this paper, we answer the following \textbf{research questions}.

\begin{enumerate}[label={(RQ~\arabic*)}, leftmargin=3em]
    \item \textit{What are the key expectations software practitioners have in a low-code tool when using it to design and develop accessible and adaptive applications?}
    \item \textit{What strategies should low-code tool developers adopt to meet the expectations of application developers?}
\end{enumerate}

In answering these questions, we provide the low-code tool development and general software engineering community with the following \textbf{research contributions}.

\begin{enumerate}[label={(RC~\arabic*)}, leftmargin=3em]  
    \item \textit{Identification of a set of developer expectations for low-code tools across all stages of the software development lifecycle.}
    \item \textit{In-depth exploration of the extent to which our proposed \toolname~approach meets these expectations, based on empirical insights.}
    \item \textit{A set of key recommendations to guide low-code tool developers in designing feature-rich, industry-ready solutions that support inclusive application development, grounded in the lessons from \toolname.}
\end{enumerate}

\section{Motivation}\label{motivation}

Consider a fictional technical lead-level developer named \textit{Alex}, whose team is tasked with developing a mobile e-banking application for a prominent national financial institution. A key client requirement is ensuring accessibility for all users, in line with a new government accessibility act. The previous app had received numerous complaints, particularly from older users, citing issues such as poor text readability, low colour contrast, overly technical language, and the removal of features like background customisation \cite{wickramathilaka2025guidelines}. This feedback resonated with Alex, who recalled helping their grandparents with tasks such as adjusting text size and resolving error messages \cite{wickramathilaka2025guidelines}.

Thus, Alex convened a team meeting to discuss how best to meet the project's accessibility requirements. They concluded that the diversity of user needs could not be adequately captured by requirements engineers within the available time and resources \cite{wickramathilaka2023, wickramathilaka2025guidelines}. Most team members acknowledged that while current practices such as applying accessibility standards, limited user testing, creating a few customised app ‘flavours’, and offering basic UI configurations were useful, they were ultimately insufficient to fulfil the client’s expectations. Although the majority supported this view, one senior developer maintained that accessibility should be the responsibility of their UX engineer:


\begin{quote}
    \textit{"I mean, I get the issues, but it's more like, [it's] not the developer's kind of aspect. It's more like a UX UI guy should solve all of this stuff, right?"} -- (A quote from developer participant P14)
\end{quote}

Alex was quick to respond that assigning sole responsibility for accessibility design to one individual is not a practical solution given the complexity and scale of the requirement.


The above fictional scenario is based on insights from an interview study described later in this paper. It highlights that current development practices are often inadequate when building accessible, personalised applications for diverse user groups such as older adults. For instance, solely relying on accessibility standards typically results in a one-size-fits-all UI that compromises individual needs \cite{wickramathilaka2025technical, wickramathilaka2023}, while manually creating tailored app ‘flavours’ is both time and cost-inefficient \cite{connor2017, wickramathilaka2025technical}.

Hence, automated approaches are essential for enabling developers to generate tailored applications that address complex accessibility and personalisation needs. These methods significantly reduce the time and resources required to create user-specific app variations compared to manual development. One promising solution is generative AI, used as a low-code tool to generate adapted app instances based on user-specific context prompts. However, concerns persist around code maintainability, as the advantages of automation diminish when developers face readability issues while reviewing or modifying the AI-generated codebase. Several interview participants questioned whether current generative AI tools are equipped to reliably handle such a complex code-generation scenario. As one participant (P18) remarked:

\begin{quote}
    \textit{"I wouldn't want generative AI to mangle with production-ready code. That is not ideal. We're not there yet."}
\end{quote}

A more pragmatic low-code approach involves leveraging traditional model-driven engineering (MDE) techniques. In this context, domain-specific languages (DSLs) can be used to capture the diverse accessibility needs of seniors and other user groups. These DSLs then serve as primary artefacts in an MDE-based code generation workflow to produce tailored app instances that address specific user requirements \cite{yigitbas2020,wickramathilaka2025technical,bendaly2018}.

At present, only a limited number of approaches exist to support the model-driven engineering (MDE) of accessible and adaptive user interfaces (e.g.,~\cite{yigitbas2020, bendaly2018, ghaibi2017, bacha2011, akiki2016, minon2015, bongratz2012}), and those that do are primarily constrained to research prototypes~\cite{wickramathilaka2023}. This gap presents a significant opportunity for tool developers to design and implement robust MDE-based solutions that facilitate the creation of accessible and adaptive applications. The importance of such tools is further underscored by emerging regulatory developments, which introduce legal and ethical imperatives for software development organisations to adopt inclusive design practices.
For example, the European Accessibility Act (EAA)~\cite{eu2019directive882} covers software products across a range of devices (e.g., smartphones and computers) and service domains (e.g., banking and transport), aiming to reduce barriers faced by individuals with disabilities in accessing products and services within EU member states. In parallel, other global regulatory instruments reinforce similar principles. These include the United Nations' \textit{Convention on the Rights of Persons with Disabilities} (CRPD)~\cite{crpd}, Section 508 of the U.S. Federal Electronic and Information Technology Accessibility Standards~\cite{USGovSection508}, France’s \textit{General Accessibility Framework for Administrations} (RGAA)~\cite{rgaa}, and Ontario’s \textit{Accessibility for Ontarians with Disabilities Act} (AODA)~\cite{aoda}.

To illustrate the practical impact of these regulatory frameworks on software development practices, consider a scenario where one European participant [P18] indicated that their organisation is actively exploring strategies to address these compliance requirements, despite not being legally obligated to do so due to their smaller organisational scale.

\begin{quote}
    \textit{``Because the EU put a new directive for accessibility regulations. And we do not fall under it [regulation] because our company is too small, so we don't have to follow it. But of course, because there is a more widespread acknowledgement of these issues, we are also investigating what can we do to improve the situation [of accessibility barriers]."}
\end{quote}


However, deploying a tool without a thorough understanding of its practical implications for its target users (in this case, software practitioners) risks poor adoption and limited success. We faced a similar challenge during the development of our own prototype low-code tool aimed at supporting accessibility and personalisation for seniors. To address this, we conducted an interview-based evaluation with professional developers to explore the enablers, barriers, and improvement opportunities they identified on our tool: \toolname. Drawing on these insights, we offer a set of practical recommendations for future low-code tool developers to better meet developer expectations in creating accessible and adaptive applications for diverse user groups.

\section{Our Approach}\label{methodology}

\subsection{Overview} \label{overview}

\begin{figure}
\includegraphics[width=0.9\textwidth]{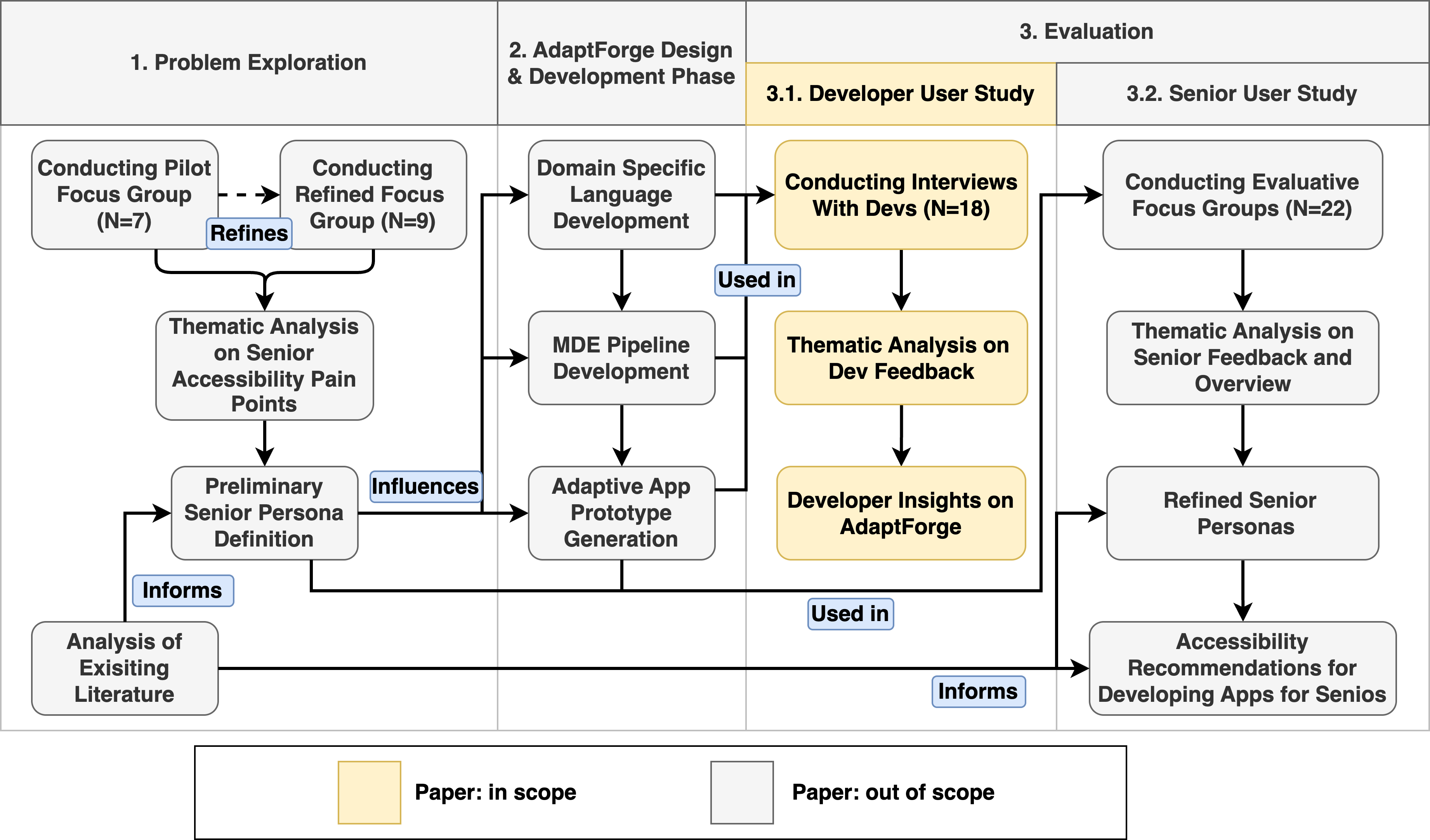}
\caption{A holistic overview of the project to design, develop, and evaluate \toolname. In this paper, we will emphasise our findings from the Developer User Study that we conducted as a part of \toolname~evaluation.}
\label{project_overview}
\Description{}
\end{figure}

\begin{figure}
\includegraphics[width=0.9\textwidth]{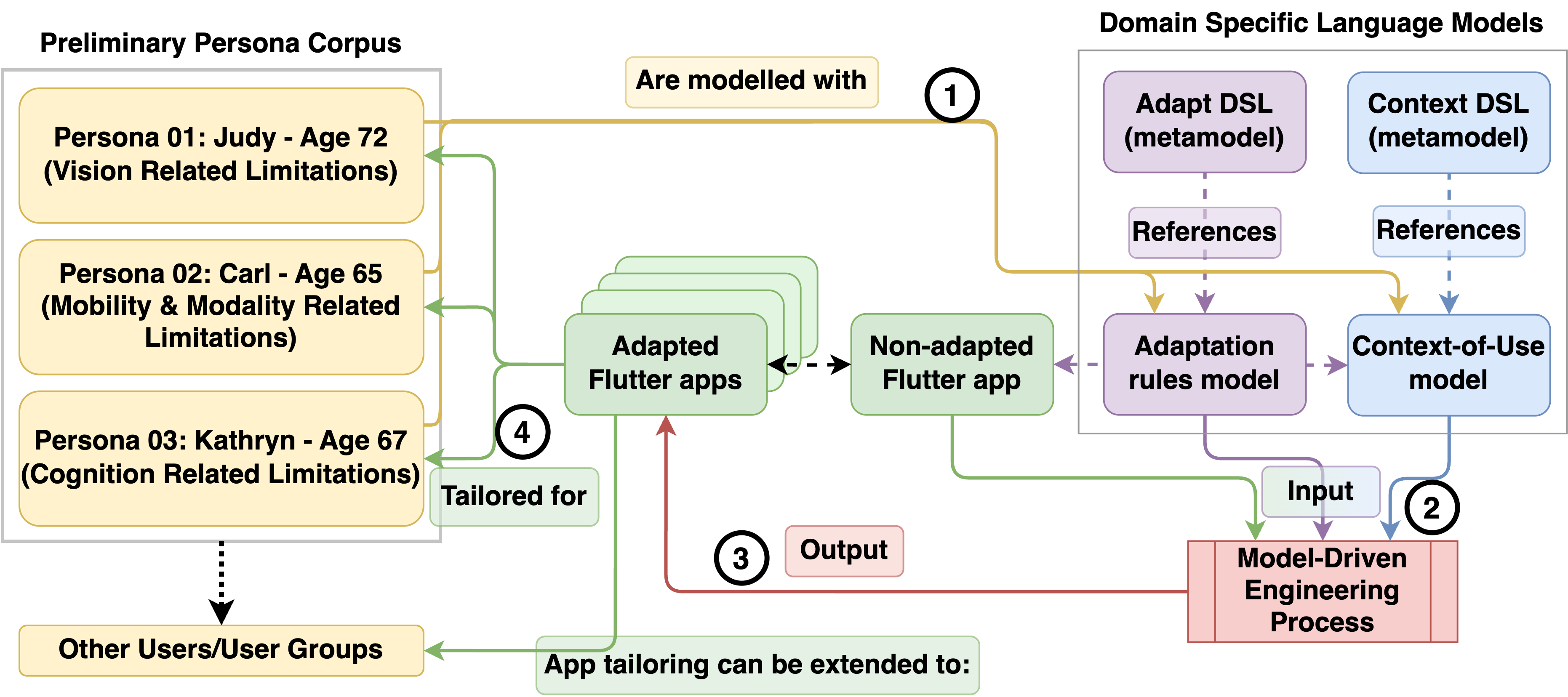}
\caption{An illustration of the the abstract architecture and workflow of \toolname.}
\label{tool_overview}
\Description{}
\end{figure}

During the initial phase of our study, we investigated the accessibility barriers faced by older adults when interacting with software applications. This investigation was conducted through a focus group study involving 16 senior participants. Our findings confirmed that older adults frequently encounter challenges related to accessibility and personalisation in their daily use of mobile applications. To contextualise these findings, we developed three preliminary personas, each representing a distinct set of age-related needs, frustrations with software, and practical workarounds employed by older adults to partially mitigate these challenges (e.g., limitations related to vision, mobility, modality, and cognition), as illustrated in Figure~\ref{tool_overview}.

Subsequently, we used these personas as design artefacts, supplemented by additional resources pertaining to accessibility and the needs of ageing users. These included relevant accessibility standards (e.g., WCAG~\cite{wcag2.2} and ISO 9241-171~\cite{iso9241}) and pertinent literature~(e.g., \cite{morey2019, ahmad2020, harte2017, watkins2014, bendaly2018, yigitbas2020}). Based on this foundation, we initiated the design and development of \toolname.

The prototype tool was evaluated from two distinct perspectives: the software practitioner/developer perspective, through an interview study involving 18 professional developers; and the senior end-user perspective, via a user study with 22 older adults from the same community involved in the exploratory stage focus groups. This continuity introduced a longitudinal dimension to the design and evaluation process.

In our earlier papers, we examined the design and development of the tool in detail~\cite{wickramathilaka2025technical}, as well as the findings from the longitudinal senior focus group studies~\cite{wickramathilaka2025guidelines}. As depicted in Figure~\ref{project_overview}, the present paper focuses on the developer user study and the insights gained from this evaluation. These insights are further generalised to inform the broader low-code software tool development community.

\subsection{Usage of \toolname} \label{tool implementation}

To describe the functionality of \toolname, we revisit the example of \textit{Alex} and their development team introduced in Section~\ref{motivation}. The team initially develops a Flutter-based cross-platform mobile application that fulfils the functional and non-functional requirements specified by the project stakeholders. Subsequently, they aim to leverage the open-source tool \toolname~to generate adapted application instances that address the accessibility and personalisation needs of senior users, as well as other disadvantaged user groups (e.g., individuals with physical or cognitive impairments).

The first step involves using the Context DSL editor within \toolname~to construct persona representations of the user base. For the initial deployment, the team focuses on three specific limitations: visual impairments, hand dexterity issues, and cognitive impairments. They create three corresponding persona profiles using the model editor, which is structured as a hierarchical tree of nodes. Each persona is defined through an $n$-tuple that captures three contextual dimensions, collectively forming a context-of-use model: (1) \textbf{User context}, encompassing impairments (e.g., visual, auditory, cognitive), privacy preferences, colour scheme preferences, language settings, and language comprehension levels; (2) \textbf{Platform context}, referring to device-specific characteristics such as screen dimensions, hardware specifications, and operating system configurations; and (3) \textbf{Environment context}, which includes external variables such as geo-location, ambient light and noise levels, and user activity patterns~\cite{wickramathilaka2025technical, calvary2002, Stephanidis2000}.
It is important to note that most environmental parameters can only be reliably detected at runtime and are thus generally unsuitable for defining design-time adaptation rules. Nevertheless, one of the goals of \toolname~is to enable these rules to operate at runtime using sensor data from the user's device to evaluate conditions and trigger adaptations. However, such runtime adaptations fall outside the current scope of this work.

Following the persona modelling, the team employs the Adapt DSL model editor. Using context-of-use information from the Context DSL, they define conditional adaptation scenarios. For instance, Alex creates a scenario using the following rule: \textit{IF userContext.visionImpairment.type == `Macular Degeneration' AND user.age \textgreater 50 AND platform.deviceDetails.name == `iPhone11'}. Once such conditions are established, they proceed to reference UI components from their existing Flutter application and define adaptation operations that are triggered upon condition satisfaction. Integration between Adapt DSL and the Flutter codebase is straightforward: developers only need to identify relevant UI widgets and annotate them with unique \texttt{Key} attributes: either individually or collectively using shared identifiers.

Once both DSL models and the Flutter application are correctly integrated, Alex triggers the MDE workflow. Within seconds, \toolname~generates three adapted versions of the original application, eliminating the need for extensive manual code modifications. Each version is tailored to meet the accessibility and personalisation requirements identified during the design phase. Alex recognises that expanding the number of app variants merely requires extending the DSL models to represents accessibility and adaptation needs of new users or groups and re-triggering the MDE workflow, potentially generating dozens or hundreds of customised application instances. 
Importantly, the original Flutter source code remains the team's source of truth. For ongoing maintenance and updates, developers only need to modify the base application and update the DSL models accordingly. The final step involves ensuring that each adapted application instance is correctly deployed to the appropriate target users.

With our current iteration of \toolname, Alex and their team can use the tool at design time to adapt any Flutter application across multiple platforms (e.g., desktop, web, native iOS, and Android), as modifications are applied directly at the Flutter source code level (a cross-platform development framework). In future iterations, the prototype will be extended through a backlog of feature improvements and additions informed by empirical insights. These enhancements span various stages of the software development lifecycle, including requirements engineering, codebase scalability during code generation, maintainability, quality assurance, and deployment processes.

With these improvements, \toolname~has the potential to become a valuable tool for supporting the development of accessible and adaptive cross-platform applications, ranging from small-scale projects managed by teams similar to Alex’s to large-scale systems consisting of hundreds of thousands of lines of code, maintained across efficient Continuous Integration and Continuous Delivery (CI/CD) workflows.

\subsection{Generation of An Adaptive Mobile Application Prototype} \label{protoype generation}

Similar to the hypothetical scenario discussed earlier, we too used our preliminary persona corpus to create adapted app instances from an open-source Flutter application\footnote{\href{https://github.com/adeeteya/FlutterFurnitureApp}{GitHub repository link for the open-source Flutter app}} from a GitHub repository and integrated it into the MDE workflow. Using our corpus of personas, we modelled the age-related accessibility needs contextualised within each persona using Context DSL and Adapt DSL.

\begin{figure}
\includegraphics[width=1\textwidth]{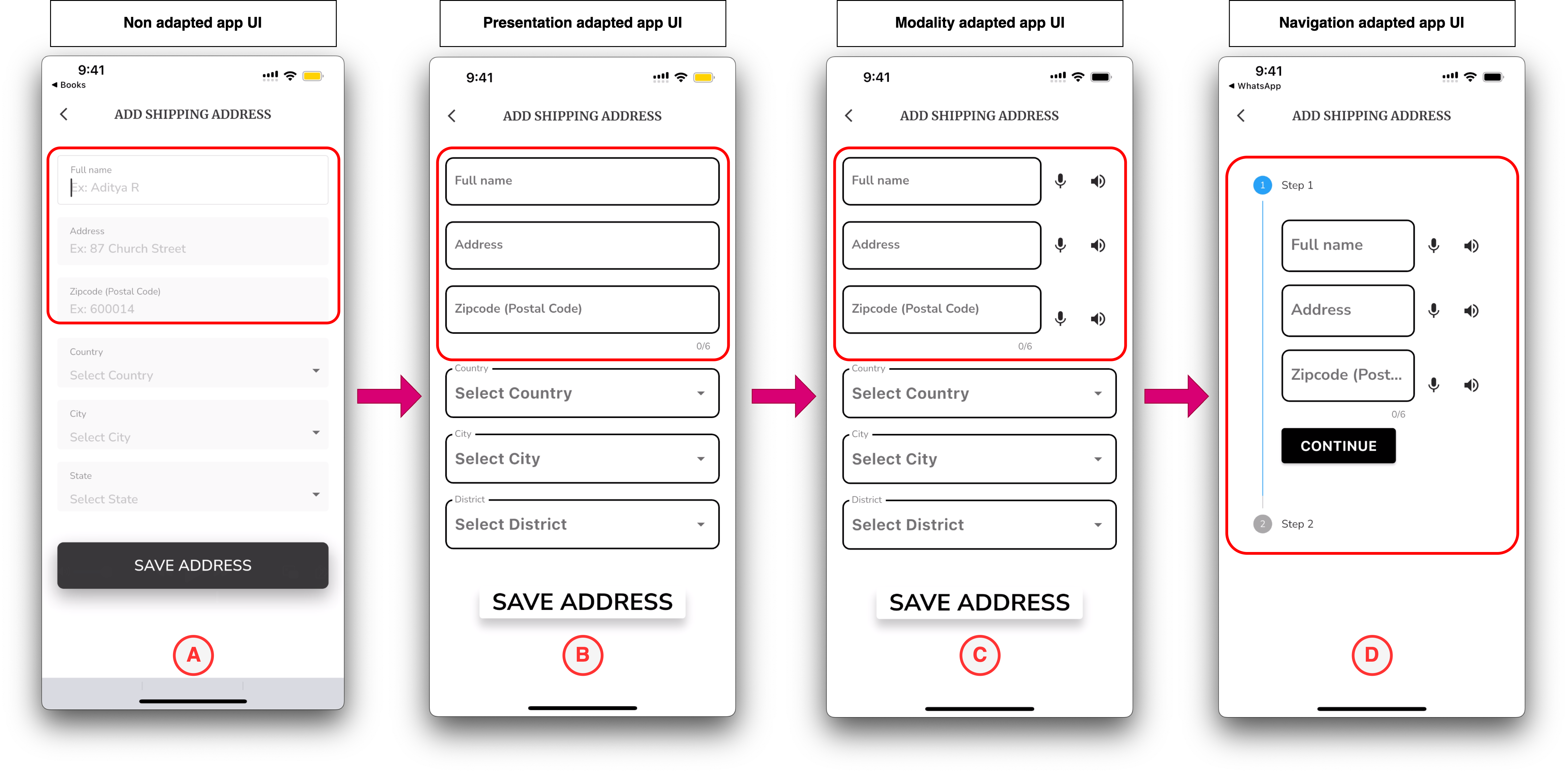}
\caption{An example of how \toolname~tool can be used to generate adapted app UI instances based on three different types of adaptations that were identified during the exploratory stage.}
\label{adaptation examples}
\Description{This figure presents an example using the \textit{Add Shipping Address} UI from an open-source furniture application. In (A), the original UI is shown. In (B), presentation-related adaptations—such as modified text input fields, dropdowns, and buttons—are applied using \toolname. In (C), the interface has been further adapted to include text-to-speech and speech-to-text functionality. Finally, (D) demonstrates a navigation-adapted version where the form is restructured into a step-wise layout to improve usability for seniors.}
\end{figure}


The first persona represented an older adult experiencing vision impairment. Adaptations for this persona focused on presentation-related modifications such as increasing text size, applying a high-contrast black-and-white colour theme, and enhancing input field borders to improve visibility (Section B, Figure \ref{adaptation examples}). The second persona illustrated challenges associated with reduced hand dexterity, often caused by conditions such as arthritis. To accommodate these needs, the adapted interface integrated multimodal input and output support, including speech-to-text for text input and text-to-speech for output, thereby reducing reliance on conventional touchscreen interaction (Section C, Figure \ref{adaptation examples}). The third persona was based on cognitive limitations, such as age-related mild cognitive impairment (MCI), which can lead to confusion and disorientation when navigating complex application menus or when unexpected UI changes occur. Adaptations for this persona focused on simplifying navigation structures and implementing step-wise forms to reduce cognitive load; These changes also helped to alleviate interface limitations arising from the enlarged UI components used in other adaptations (Section D, Figure \ref{adaptation examples}).

\section{Developer Interview Study}


With a working prototype of \toolname, we aimed to determine its reception among the two key stakeholder groups involved in our project: 1) Software practitioners/developers are the primary users of \toolname, while 2) Seniors and other end users ultimately benefit from its adoption. While we considered both perspectives during the evaluatory phase of the study, we do not revisit the end-user part of the evaluation as it is already comprehensively discussed in Wickramathilaka et al.~\cite{wickramathilaka2025guidelines}.

Instead, this paper focuses on the developer interview study, which involved 18 professional software developers participating in a qualitative, interview-based investigation. The study yielded a wide range of insights that are not only applicable to the refinement of \toolname~but can also be generalised to inform the broader low-code development community. Specifically, the insights reflect participants’ perspectives on how a low-code tool, designed for software practitioners to build more accessible and adaptive applications, can be integrated across various phases of the software development life cycle, offering significant automation benefits to development teams.

Accordingly, in this section, we first present the study protocol and details of the participant cohort. This later serves as a foundation for the generalised recommendations provided in subsequent sections.

\subsection{\textbf{Study Design:}}

To begin with, we obtained ethical approval from the Monash Human Research Ethics Committee (MUHREC Project ID: 42470) to conduct this study. Given that \toolname~was developed using the Eclipse Modelling Framework \cite{emf} and several associated plugins such as Xtext \cite{xtend}, Sirius \cite{sirius}, and Acceleo \cite{acceleo}, expecting participants to install the tool, configure its dependencies within the Eclipse environment, use it, and subsequently participate in an interview or survey would have posed a significant time burden, particularly for professional practitioners with limited availability. Therefore, we adopted a more pragmatic, video-based approach. We created a screen-recorded demonstration\footnote{\href{https://drive.google.com/drive/folders/11b_P07ybFD7SoDS1r7rrdWd7wks1la6n?usp=sharing}{Google Drive link for video demonstrations}} in which one of the authors used \toolname~to address the accessibility requirements of a synthetic persona that consolidated the age-related needs represented across all three original personas used in the tool’s design. This method allowed us to present the core functionality of the tool in a concise and comprehensible manner while significantly reducing the time commitment required from participants, without compromising the richness or relevance of the data collected.

Each interview session was scheduled for approximately one hour. At the beginning of the session, participants were introduced to a hypothetical scenario in which they were involved in a software development project targeted at an application with a significant number of senior users aged 60 and above. A brief explanation was then provided regarding how ageing contributes to diverse and heterogeneous accessibility needs among older adults, along with a short overview of how \toolname~seeks to address these needs by generating adapted app instances tailored to individual senior users.

The first part of the video demonstration introduced the accessibility needs of the condensed persona. These needs were then used to demonstrate how the persona could be modelled using the \textbf{context DSL} editor to create a \textit{context-of-use} model instance. Following this demonstration, participants were asked to evaluate the practicality and usefulness of the context DSL tool, its ease of use, suggestions for improvement, and its applicability in real-world software development projects.

Building upon the same scenario, the next demonstration involved using the \textbf{Adapt DSL} editor to define conditional adaptation rules that specify the required accessibility adaptations and indicate where in the Flutter app codebase these should be applied. Subsequently, the process of feeding the DSL models and the app source code into the \toolname~Model-Driven Engineering pipeline was shown, resulting in the generation of an adapted Flutter app instance. As with the previous step, participants were asked to reflect on the usefulness, usability, and practical feasibility of the Adapt DSL tool and to provide any suggestions for improvement, including its potential for integration in professional practice.

The final segment of the demonstration showcased the application of more complex adaptations to the app codebase. These adaptations went beyond basic presentation modifications (e.g., increasing text size or boldness) that were demonstrated earlier and included the integration of features such as text-to-speech and speech-to-text capabilities, as well as the transformation of a traditional form interface into a dynamically grouped, step-wise wizard-style form. Participants were again invited to comment on the usability, usefulness, and real-world relevance of these capabilities.

At the conclusion of each interview, participants were asked to provide their overall impressions of \toolname~and to share whether they would consider adopting such a tool in their own development workflows.

\subsection{\textbf{Data Collection and Analysis:}}

All 18 interview sessions were conducted by one of the authors. With the exception of a single in-person session, all interviews were carried out via Zoom video conferencing software. Audio recordings of the sessions were captured using Zoom and subsequently uploaded to the Otter.ai web-based transcription tool to generate textual transcripts. Each transcript was then manually cleaned to correct transcription inaccuracies and to ensure the anonymity of participants.

The cleaned transcripts were then subjected to a thematic analysis. During the initial coding phase, we familiarised ourselves with the dataset and generated a diverse set of codes covering various aspects of \toolname. These initial codes were further refined through multiple iterations, resulting in a hierarchical structure of themes. The top-level themes identified were as follows: (1) Positive Reactions from Developers towards \toolname, (2) Negative Reactions from Developers towards \toolname, (3) Suggestions for Improving \toolname, and (4) Other Related Themes. This structure reflects the evaluative nature of the study. Each main theme encompassed multiple sub-themes, as participants provided detailed insights spanning technical, operational, and experiential dimensions relevant to the problem space. 

For instance, the theme \textit{Suggestions for Improvement} exhibited the highest degree of variability, as participant feedback was influenced by a range of factors such as industry experience, front-end versus back-end development expertise, types of products developed, and organisational practices (e.g., Agile versus Waterfall development methodologies).

After the evaluative theme identification, we reinterpreted the key sub-themes identified across all of the top-level themes in order to understand the expectations the participants had of our tool based on their experiences as software practitioners. In this paper, we primarily report our analysis based on this reinterpreted theme structure.

\subsection{\textbf{Participant Information:}} \label{participant information}

\begin{table}
    \centering
    \renewcommand{\arraystretch}{1.2} 
    \resizebox{1\textwidth}{!}{
    \setlength{\tabcolsep}{5pt} 
    \small 
    \begin{tabular}{|l|>{\raggedright\arraybackslash}p{1.5cm}|>{\raggedright\arraybackslash}p{1cm}|>{\raggedright\arraybackslash}p{1.6cm}|>{\raggedright\arraybackslash}p{1.6cm}|>{\raggedright\arraybackslash}p{1.6cm}|>{\raggedright\arraybackslash}p{2cm}|>{\raggedright\arraybackslash}p{3cm}|}
        \hline
        \textbf{ID} & \textbf{Developer profile taxonomy} & \textbf{Age} & \textbf{Software industry exp.} & \textbf{Mobile dev. Exp.} & \textbf{Flutter exp.} & \textbf{Use accessibility standards?} & \textbf{Importance of addressing age-specific UI needs} \\ \hline\hline
        P1  & Pragmatist & 21 - 34  & 5 - 10 years  & 2 - 4 years  & 2 - 3 years  & Yes    & Important  \\ \hline
        P2  & Advocate & 21 - 34  & 5 - 10 years  & \texttt{<} 1 year  & \texttt{<} 1 year  & Yes    & Essential  \\ \hline
        P3  & Advocate & 21 - 34  & 2 - 4 years   & \texttt{<} 1 year  & \texttt{<} 1 year  & Yes    & Essential  \\ \hline
        P4  & Sceptic & 21 - 34  & 5 - 10 years  & 5 - 10 years  & 2 - 3 years  & Yes    & Essential  \\ \hline
        P5  & Advocate & 21 - 34  & 2 - 4 years   & \texttt{<} 1 year  & \texttt{<} 1 year  & Yes    & Essential  \\ \hline
        P6  & Pragmatist & 21 - 34  & 5 - 10 years  & 5 - 10 years  & \texttt{>} 5 years  & Yes    & Important  \\ \hline
        P7  & Pragmatist & 21 - 34  & 5 - 10 years  & 2 - 4 years  & 2 - 3 years  & Maybe  & Important  \\ \hline
        P8  & Advocate & 21 - 34  & 2 - 4 years   & 2 - 4 years  & 2 - 3 years  & Maybe  & Important  \\ \hline
        P9  & Pragmatist & 21 - 34  & \texttt{<} 1 year  & \texttt{<} 1 year  & \texttt{<} 1 year  & Yes    & Essential  \\ \hline
        P10 & Sceptic & 21 - 34  & 5 - 10 years  & 5 - 10 years  & \texttt{>} 5 years  & Maybe  & Important  \\ \hline
        P11 & Advocate & 21 - 34  & 2 - 4 years   & 2 - 4 years  & 3 - 4 years  & Maybe  & Essential  \\ \hline
        P12 & Advocate & 35 - 59  & \texttt{>} 10 years    & 2 - 4 years  & 3 - 4 years  & Yes    & Important  \\ \hline
        P13 & pragmatist & 21 - 34  & \texttt{<} 1 year  & \texttt{<} 1 year  & \texttt{<} 1 year  & No     & Somewhat Important  \\ \hline
        P14 & Sceptic & 21 - 34  & 5 - 10 years  & 5 - 10 years  & 3 - 4 years  & No     & Important  \\ \hline
        P15 & Pragmatist & 35 - 59  & \texttt{>} 10 years    & \texttt{>} 10 years   & \texttt{>} 5 years   & Yes    & Essential  \\ \hline
        P16 & pragmatist & 35 - 59  & 5 - 10 years  & 5 - 10 years  & 2 - 3 years  & Maybe  & Somewhat Important  \\ \hline
        P17 & Advocate & 21 - 34  & 5 - 10 years  & 5 - 10 years  & 3 - 4 years  & Yes    & Somewhat Important  \\ \hline
        P18 & Advocate & 21 - 34  & \texttt{>} 10 years  & 2 - 4 years  & 3 - 4 years  & No & Somewhat Important  \\ \hline
    \end{tabular}
    }
    \caption{Demographic data of software developer participants. The \textit{Developer profile taxonomy} field summarises each participant's general attitudes toward adopting tools aimed at enhancing accessibility-related user experiences. \textit{Advocates} are those who actively support the use of such tools, provided they improve end-user experience. \textit{Pragmatists} adopt new tools only if they align well with current workflows and reduce development effort. \textit{Sceptics} are the most resistant to change, preferring to continue using familiar tools and practices rather than adopting new ones.}
    \label{tab:developer_data}
\end{table}

We recruited 18 participants ($N=18$) from various global regions for this study. One-third of the participants were located in Oceania ($n=6$), and another third were based in Asia ($n=6$). The remaining participants were situated in Europe ($n=4$), North America ($n=1$), and Africa ($n=1$). Recruitment was conducted through social media platforms, primarily Reddit and LinkedIn, as well as via the authors' personal and professional networks. All participants were volunteers and received no financial or material incentives for participation, aside from an intrinsic interest in the objectives of the study. The primary recruitment criterion was that participants must have professional experience in software application development, and they should have a minimum of six months of professional experience with the Flutter framework. 

The demographic data that we collected during our developer interview study is provided in Table \ref{tab:developer_data}. As for specific data attributes, firstly, we inquired about participants on their age group. The sole reason for this was to understand if the age of a developer impacts their perceptions on designing and developing apps for seniors. However, only a minority (16.7\%, $n=3$) of the participants fell within the 35-59 age category, while the rest (83.3\%, $n=15$) were between 21 - 34 years, indicating a relatively young population sample of developers.

Thereafter, we assessed participants’ software development experience across three key dimensions: (1) overall professional experience as developers, (2) experience in mobile development, and (3) experience with the Flutter framework. Regarding the first dimension, half the sample (50\%, $n=9$) reported having 5 - 10 years of professional experience. Additionally, three participants (16.7\%) had over 10 years of experience, with one of them ([P15]) specifically noting during the interview that they possessed more than 30 years of experience in software development. Among the remaining participants, two (11.1\%) were novice developers with less than one year of total industry experience, while the rest (22.2\%, $n=4$) reported having 2 - 4 years of experience.

With respect to mobile application development, participants reported varying levels of proficiency. One-third ($n=6$) had been involved in the field for 5 - 10 years, while an equal number ($n=6$) reported 2 - 4 years of experience. Only one participant ([P15]) indicated more than a decade of experience in mobile development. The remaining participants (27.8\%, $n=5$) had been working in this area for less than one year.

Participants’ familiarity with the Flutter framework showed a pattern similar to that observed in mobile application development. Comparable proportions of the sample (27.8\%, $n=5$) reported using Flutter for 3 - 4 years, 2 - 3 years, and less than one year, respectively. Notably, three participants (16.7\%) indicated over five years of engagement with the framework. This is particularly noteworthy considering that Flutter 1.0 was officially released only in late 2018 \cite{flutter}.

Next, we sought to understand the current practices and strategies employed by participants when developing accessible software, particularly in contexts involving older adult users. An encouraging insight was that a majority of participants (55.6\%, $n=10$) reported incorporating accessibility standards, such as the Web Content Accessibility Guidelines (WCAG), during app development. Among the remainder, a subset (27.8\%, $n=5$) were uncertain about their use of such guidelines, while the rest (16.7\%, $n=3$) indicated that they do not use them.

We also assessed participants’ perceptions of the importance of addressing age-related needs in user interface (UI) design and development. In response to a Likert scale-based question, seven participants (38.9\%) considered this issue to be \textit{essential}, while another seven (38.9\%) rated it as \textit{important}, indicating a generally positive outlook toward the importance of inclusive design for seniors. The remaining participants (22.2\%, $n=4$) viewed it as \textit{somewhat important}. Notably, no participants rated these considerations as unimportant in the context of app development.

However, these quantitative insights alone do not provide a complete understanding of the participants' willingness to adopt a low-code tool aimed at enhancing accessibility and personalisation aspects of applications. For instance, [P4]: a user experience engineer who actively applies accessibility guidelines and various strategies to enhance user experience, expressed resistance toward adopting new tools and prioritising end-user needs over visual identity and UI uniformity of an app (i.e., prioritising client requirements over user needs). In contrast, [P8]: a software engineer with limited prior engagement in accessibility or personalisation in their development practices, expressed significant enthusiasm for adopting a low-code tool that facilitates the creation of improved user experiences for all users.

To explore these differences in greater depth, qualitative insights from the interviews were analysed to construct a taxonomy based on participants’ willingness to adopt a low-code tool for creating accessible and adaptive applications. Within this classification, only a minority (16.7\%, $n=3$) were identified as \textit{sceptics}. Among them, [P14] demonstrated consistent scepticism throughout the session, whereas [P10] transitioned from a sceptical stance to a more reserved pragmatist position during the course of the interview.

The largest subgroup (44.4\%, $n=8$) was classified as \textit{pragmatists}: participants who would consider adopting a low-code tool, provided it aligns with existing workflows, does not introduce undue complexity, and delivers clear, tangible benefits to their app development process. The remaining participants (38.9\%, $n=7$) were classified as \textit{advocates}, who prioritise enhancing end-user experience and are willing to champion the adoption of tools that enable the development of more accessible and personalised applications.

\begin{table}
    \centering
    \renewcommand{\arraystretch}{1.3} 
    \resizebox{\textwidth}{!}{ 
    \begin{tabular}{|p{6cm}|*{18}{>{\centering\arraybackslash}p{0.2cm}|}p{0.7cm}|}
        \hline
        \textbf{Personalisation Approach} & \textbf{P\newline1} & \textbf{P\newline2} & \textbf{P\newline3} & \textbf{P\newline4} & \textbf{P\newline5} & \textbf{P\newline6} & \textbf{P\newline7} & \textbf{P\newline8} & \textbf{P\newline9} & \textbf{P\newline10} & \textbf{P\newline11} & \textbf{P\newline12} & \textbf{P\newline13} & \textbf{P\newline14} & \textbf{P\newline15} & \textbf{P\newline16} & \textbf{P\newline17} & \textbf{P\newline18} & \textbf{\%} \\ \hline
        
        Design a universally accessible interface based on accessibility guidelines & \cellcolor{green!50}
         & \cellcolor{green!50} & \cellcolor{green!50} & \cellcolor{green!50} &  & \cellcolor{green!50} & \cellcolor{green!50} &  &  &  & \cellcolor{green!50} &  & \cellcolor{green!50} & \cellcolor{green!50} & \cellcolor{green!50} &  & \cellcolor{green!50} & \cellcolor{green!50} & \textbf{66.7\%\newline(12)} \\ \hline
        
        Ensure UI integrates with native accessibility settings &  & \cellcolor{green!50} & \cellcolor{green!50} & \cellcolor{green!50} & \cellcolor{green!50} & \cellcolor{green!50} & \cellcolor{green!50} &  &  &  &  & \cellcolor{green!50} &  & \cellcolor{green!50} & \cellcolor{green!50} & \cellcolor{green!50} &  & \cellcolor{green!50} & \textbf{61.1\%\newline(11)} \\ \hline
        
        Capture personalisation needs through user interviews and personas & \cellcolor{green!50} & \cellcolor{green!50} & \cellcolor{green!50} &  & \cellcolor{green!50} &  &  &  & \cellcolor{green!50} &  &  &  & \cellcolor{green!50} &  & \cellcolor{green!50} &  &  &  & \textbf{38.9\%\newline(7)} \\ \hline
        
        Provide a configuration dashboard for UI adjustments & \cellcolor{green!50} &  &  & \cellcolor{green!50} & \cellcolor{green!50} & \cellcolor{green!50} &  &  & \cellcolor{green!50} &  &  &  &  & \cellcolor{green!50} & \cellcolor{green!50} &  &  &  & \textbf{38.9\%\newline(7)} \\ \hline
        
        Special case handling: Stick with general practices unless specific requirements exist &  &  &  &  &  &  &  &  &  &  &  &  &  &  &  &  & \cellcolor{green!50} &  & \textbf{5.6\%\newline(1)} \\ \hline
        
        Rarely take steps to personalise UI due to time/resource constraints &  &  & \cellcolor{green!50} &  &  &  &  & \cellcolor{green!50} &  & \cellcolor{green!50} &  &  &  &  &  & \cellcolor{green!50} &  &  & \textbf{22.2\%\newline(4)} \\ \hline
    \end{tabular}
    }
    \caption{Current strategies used by developer participants to address age-related UI needs}
    \label{tab:personalisation_strategies}
\end{table}

The final question in our demographic data collection survey (Table~\ref{tab:personalisation_strategies}) aimed to elicit detailed insights into the strategies software developers employ when designing user interfaces and applications for older adults. The most frequently reported approach was adherence to established accessibility standards and guidelines, such as WCAG~\cite{wcag2.2} and ISO~9241-171~\cite{iso9241}, to create universally accessible application interfaces (66.7\%, $n=12$). This was closely followed by compliance with the accessibility requirements, guidelines, or settings prescribed by the native development platform (61.1\%, $n=11$). Regarding personalisation strategies, a smaller proportion of participants reported using qualitative user research methods such as user interviews and persona development, to identify the specific needs of older users (38.9\%, $n=7$). An equal proportion (38.9\%, $n=7$) indicated the provision of configuration dashboards that allow users to tailor interface settings according to their preferences. Conversely, a minority (22.2\%, $n=4$) acknowledged infrequent implementation of UI personalisation for older adults, citing time and resource constraints as primary barriers.

\section{Results}\label{results}


\subsection{Dev Expectation 01: The Need to Fulfil End-User Accessibility-Adaptation Requirements}

\subsubsection{\textbf{Context:}}

During our interview study, a key positive reaction that we received from a majority of participants ($n=13$) was the recognition of how the \toolname~can enhance the end-user experience. For example, participant [P10] expressed a willingness to advocate for the adoption of our \textit{context DSL} modelling tool to their clients, even in the face of trade-offs.

\begin{quote}
    \textit{"I think it's worth convincing the client regarding these stuff [modelling contextual parameters of users], because it really [helps to] improve the user experience."}
\end{quote}

Such statements led us to further investigate what motivated participants to adopt a user-centric approach when addressing the accessibility and personalisation needs of their user base. We aimed to understand whether this inclination was driven by personal empathy toward the accessibility barriers faced by end-users, or if it was primarily influenced by professional responsibilities and expectations within their roles.

\subsubsection{\textbf{Reality:}}

We identified explicit references from 11 participants who acknowledged the challenges faced by seniors due to age-related limitations. In seven of these cases, participants shared personal anecdotes involving an older family member, such as a parent or grandparent, who had encountered accessibility barriers. A few participants ($n=7$) also mentioned that they themselves had begun to experience similar limitations. Overall, 15 out of 18 participants demonstrated a user-centred mindset at least once during the interview sessions by considering the perspective of end-users affected by impairments that compromise their user experiences. These insights suggest that software practitioners are leaning more towards personal empathy for the needs of their user base rather than acting out of mere professional obligation.

Furthermore, beyond personal motivation, institutional factors may also influence developers to adopt tools that support accessibility. For instance, participant [P18] highlighted a growing emphasis on accessibility within the European Union, where regulatory developments are pushing companies toward more inclusive design practices. The following comment from [P18] captures this sentiment:

\begin{quote}
    \textit{"I think it [\toolname] has a bigger market now that, especially in the EU, I mean, we have like a couple 100 million people living here, and all bigger companies will have to adapt accessibility paradigms now effectively. So they have to look into solutions, because most of them don't."}
\end{quote}

\subsubsection{\textbf{Potential Solutions:}}

Based on the positive feedback received from our participants, we believe that \toolname~aligns well with the expectations of the majority ($n=17$), serving as a complementary development tool that supports the creation of more accessible and personalised applications tailored to the diverse needs of older adults. For example, participant [P8] noted:

\begin{quote}
    \textit{"I don't actually see a reason for a developer not to use it [\toolname] if it is publicly available because, if it can do that much of complex adaptations on the go, everybody would like to use it. Because we as developers, we want to give the user the best experience that we can give, and at the same time, we want more users to use our application. So we would be actually neglecting, like around 20\% 30\% of our audience because of these accessible related issues. So my overall impression is, it is something I would like to actually try on in [developing an] application in the near future."}
\end{quote}

While the current prototype was widely praised, participants also provided constructive criticism. However, almost all of these comments were largely intended to improve the tool’s viability as a software product, rather than rejecting the underlying concept. Several participants even envisioned \toolname~as an open-source or commercial tool/package ([P3], [P6], [P8], [P12], [P18]), with two participants ([P6] and [P12]) offering to contribute as open-source collaborators or to explore funding opportunities to help bring the product to market.

\subsection{Dev Expectation 02: Easy Collection and Use of Accessibility Data}

\subsubsection{\textbf{Context:}}

With one of our domain-specific languages, \textit{context DSL}, we aim to assist developers in capturing and mapping user, platform, and environmental context parameters that may influence the application experience of senior users. However, in its current iteration, the design and functionality of \textit{context DSL} were prioritised over its practical usability. Therefore, one of the objectives of our interview study was to explore the practical implications and expectations of developers when collecting and modelling contextual parameters of end-users.

\subsubsection{\textbf{Reality:}}

Several participants ($n=5$) raised concerns about the practical difficulties involved in collecting various context parameters from end-users. The conventional method, which involves requesting users to input this information manually, typically during onboarding, can negatively impact the user experience. This issue is particularly significant for older adults, who may have age-related impairments that make typing or navigating interfaces challenging. According to [P3] and [P4], such usability obstacles could lead to a reduction in the application's user base. [P2] summarised this concern succinctly:

\begin{quote}
    \textit{"So I think if user has to input those information themselves, especially when they are older, at some point, they are gonna be super annoyed by this."}
\end{quote}

Even when developers are able to obtain the relevant data, they may be unwilling to manually define context-of-use models for each individual user. As noted by three participants, this process is both time-consuming and repetitive, reducing the feasibility of such an approach in scalable development environments.

\subsubsection{\textbf{Potential Solutions:}}

To address the challenge of collecting user data, participants proposed several strategies. The simplest approach, as suggested by [P2] and [P18], is to limit the amount of information collected during onboarding. For instance, the process can prioritise obtaining essential details such as age and impairments. If users require more granular customisations, they can voluntarily provide additional information via a dedicated context information dashboard. Crucially, it is important to clearly communicate the purpose of data collection and how the information will be used, to alleviate any concerns about privacy violations.

Another proposal, raised by [P8], [P12], and [P15], involves establishing a universal platform where users can enter their context parameters with the understanding that the data will be used to personalise their application experiences. Such an accessibility information platform could offer substantial benefits, particularly for developers aiming to support consistent user experiences across multiple applications. This is especially valuable in institutional contexts, such as government services. The following example from [P12] illustrates this concept:

\begin{quote}
    \textit{"Hey, look, you can create yourself a profile of your impairments over here. And then apps can subscribe to that service so that when that person, when Judy, logs into this new app, she needs to be able to [use] the New Zealand travel declaration, the New Zealand traveller declaration [app] will go over to that service, grab her profile and then apply it all to the UI, yeah? So she doesn't have to do that for every single app."}
\end{quote}


To address the additional modelling effort required from developers, both aforementioned strategies offer potential mitigation. Information voluntarily provided by end-users can be structured in formats such as XML or JSON, enabling direct transformation into context-of-use models without necessitating manual developer input. Furthermore, as noted by four participants ([P2], [P4], [P13], and [P15]), developers may opt to model user groups rather than individual users using our DSLs. This practice reduces the number of application instances that the model-driven engineering process must generate. This approach presents a practical solution for organisations to deliver meaningful personalisation while avoiding the complexity associated with maintaining numerous application variants.

\subsection{Dev Expectation 03: Ease of Use in Modelling Tools}

\subsubsection{\textbf{Context:}}

Adopting a new tool, such as a modelling framework, requires developers to first become acquainted with its syntax and modelling grammar. We considered this in the design of our DSLs, aiming to ensure that the proposed domain-specific languages are intuitive and accessible to software practitioners, thereby reducing the barrier to adoption. During our user study, participants responded rather positively to the usability of the modelling tools. However, a more detailed analysis of their feedback is needed to identify the specific features that contributed to meeting developers' usability expectations.

\subsubsection{\textbf{Reality:}}

\begin{figure}
\includegraphics[width=0.7\textwidth]{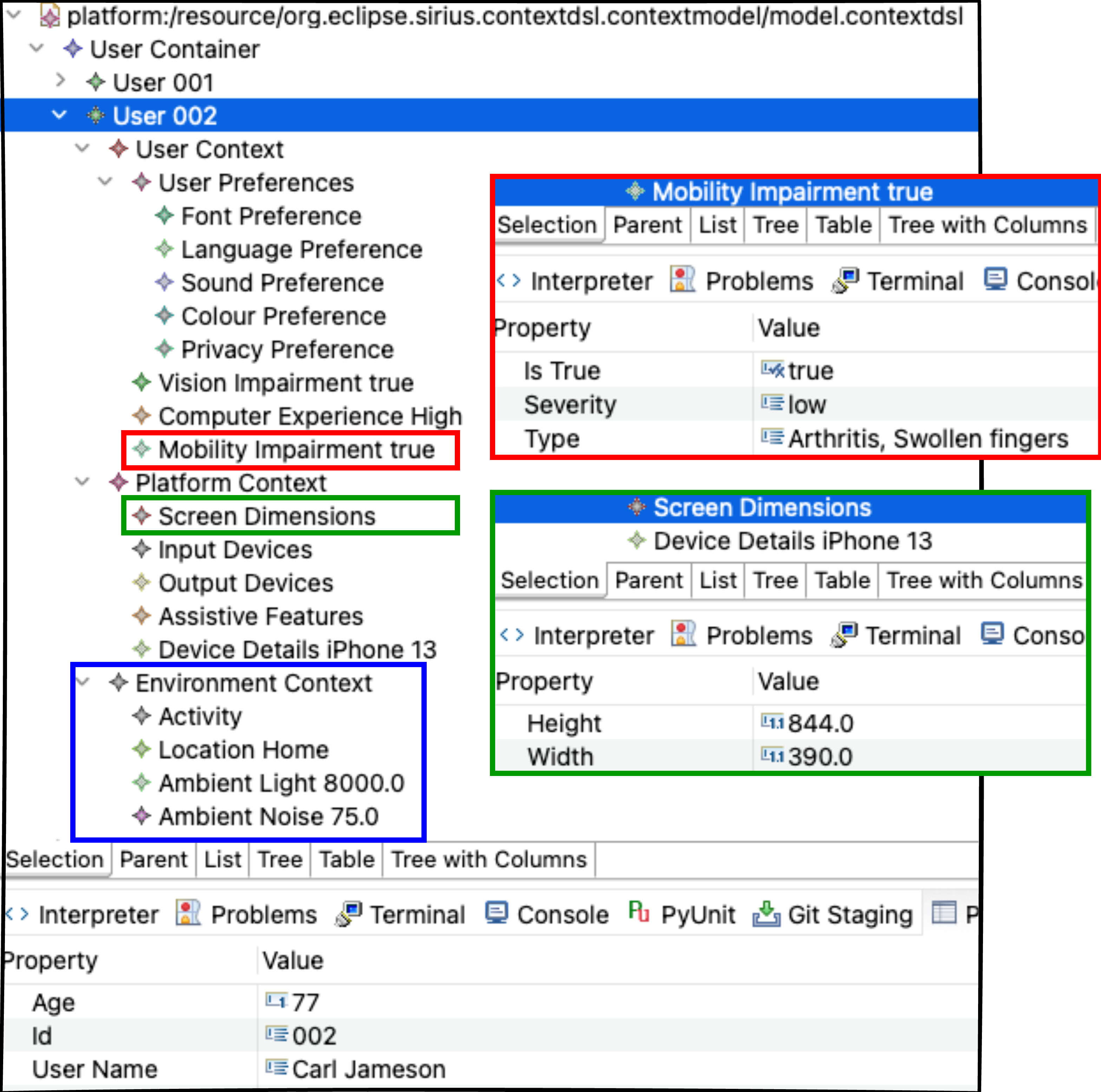}
\caption{An example Context-of-Use model defined using the context DSL editor. It captures the user, platform, and environmental context parameters for a 77-year-old senior named Carl Jameson. The model follows a tree-node hierarchy, allowing the developer to insert sub-nodes under the three main categories. When a node is selected, the \textit{Properties} view displays the associated attributes. For instance, the \textit{Mobility impairment} parameter under the \textit{User context} is highlighted in red, the property view for \textit{Screen dimensions} under the \textit{Platform context} is highlighted in green, and several example nodes under the \textit{Environment context} are highlighted in blue.}
\label{contextdsl_example}
\Description{}
\end{figure}

\textit{context DSL}, as shown in Figure~\ref{contextdsl_example}, is a semi-graphical modelling language. Of the 15 participants who provided positive feedback on its user-friendliness, 9 specifically highlighted the \textit{structure} of the DSL. They appreciated its tree-node hierarchy, which allows developers to define various context-of-use parameters as sub-nodes (and sub-sub-nodes, if needed) under three primary parent nodes: \textit{User context}, \textit{Platform context}, and \textit{Environmental context}. This structure was perceived as both familiar and logical, contributing to a positive expectation regarding the tool’s intuitiveness and a gentle learning curve for practical adoption.



\begin{figure}
\includegraphics[width=0.9\textwidth]{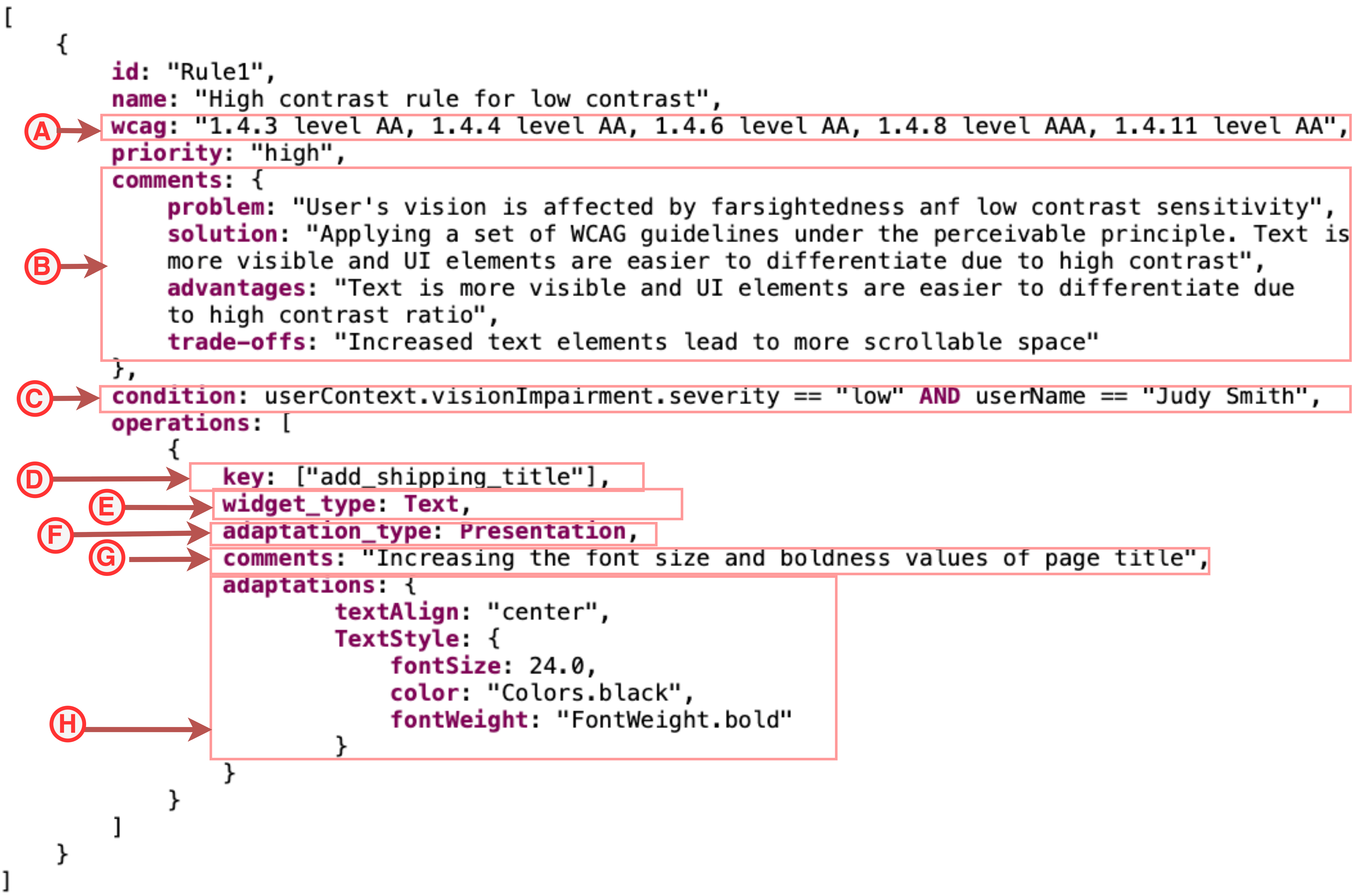}
\caption{An example adaptation rules model for a senior named Judy (a runtime model instance of Adapt DSL). To explain each section in the model: A) WCAG reference field, B) Overall comment definition of the rule, C) Context DSL referencing conditional statement, D) Unique widget(s) reference field, E) Widget type field, F) Adaptation type field, G) Comment definition field for a granular adaptation operation, and H) Example adaptation operations for a Flutter Text widget presentation adaptation.}
\label{adaptdsl_example}
\Description{}
\end{figure}

However, a tree-node structure is not the only way to design an intuitive modelling tool, as demonstrated by our alternative approach, \textit{Adapt DSL}, which also received positive feedback from 15 participants. Unlike \textit{context DSL}, \textit{Adapt DSL} is more concrete, defining adaptation rules that specify \textit{why}, \textit{where}, and \textit{how} source code should be modified based on conditional logic linked to context parameters defined in a \textit{context DSL} model. Due to this complexity, we adopted a self-descriptive, textual modelling language that aligns closely with JSON conventions (Figure~\ref{adaptdsl_example}). This design choice was well received, with 11 participants specifically appreciating its object-based structure, which resonated with their existing development practices. For example, participant [P7] highlighted how the familiarity of \textit{Adapt DSL} extended beyond its JSON-like structure:

\begin{quote}
    \textit{"I really like that you use parentheses and square brackets to mimic the objects and arrays so that it is more aligned with all the other languages, [so] that it is not something foreign for a developer."}
\end{quote}

\subsubsection{\textbf{Potential Solutions:}}

The positive feedback received suggests that participants appreciated the familiarity and logical structure conveyed by our use of conventional formats, such as a traditional tree hierarchy and JSON-style object representation. As participant [P9] noted:

\begin{quote}
    \textit{"To me, it is like we are always using that [a tree-node hierarchy]. Like IDEs, like creating folders and files, [it's the] same structure. So we are very familiar with this structure."}
\end{quote}

We also asked participants whether they would prefer a more graphical form of our DSLs and, if so, what their ideal model representation would be. Overall, enthusiasm for fully graphical domain-specific languages was limited; only participant [P17] explicitly favoured such a format. Others, including [P1] and [P3], expressed a clear preference for the current semi-graphical and textual representations. Participants [P13] and [P16] recommended minor graphical enhancements to \textit{context DSL}, such as incorporating intuitive icons within nodes to visually communicate context parameters or classifications. Participant [P1] offered an alternative perspective, proposing that graphical models might be more intuitive for non-technical roles within development teams, such as UX engineers, product owners, and domain experts, who may wish to contribute to the modelling process. In such cases, a DSL that provides both a technically detailed view for developers and a simplified, visual perspective for non-technical stakeholders could support more seamless collaboration across interdisciplinary teams.

\subsection{Dev Expectation 04: Utility of MDE Tools}

\subsubsection{\textbf{Context:}}

Due to the diverse nature of accessibility and adaptation needs among older adults, our modelling tools were designed to be broad in scope and extensible in depth, particularly in their capacity to capture user needs. This design decision proved advantageous, as it aligned with participants' expectations that the models could be readily extended to represent the needs of user groups beyond seniors.

\subsubsection{\textbf{Reality}}


While \toolname~was designed from the outset with extensibility in mind to accommodate a broad range of accessibility and personalisation needs arising from the diverse intersectional identities of end-users (e.g., variations in physical, cognitive, cultural, and linguistic characteristics), the demonstration in this user study was intentionally focused on addressing the accessibility and personalisation barriers faced by the senior community. This focus led to some confusion among participants, with several ($n=6$) expressing concern that the proposed tool should not be limited to supporting only senior users, as they could envision significantly broader applicability for \toolname. For example, [P4] noted the following:

\begin{quote}
    \textit{"But with this, we will be able to get a more personalised experience for each of our individual users, and we can attend [to] them individually. Like, this will be a much greater experience for everyone, not only senior people."}
\end{quote}

Participants also envisioned additional, unintended use cases for the tools. For example, participants [P1], [P7], and [P11] emphasised the importance of flexibility, expressing a preference for highly configurable model-driven engineering (MDE) tools, so that they have a greater freedom in adapting their app source code through DSLs. In another case, [P2] and [P13] identified potential for \toolname~as a rapid UI prototyping tool, although with differing motivations. [P2] intended to use it to generate high-fidelity prototypes for end-user and co-designer feedback, while [P13] saw value in using \toolname~as a simulation environment for evaluating application behaviour under varying platform and environmental conditions.

\subsubsection{\textbf{Potential Solutions}}


The full breadth of the unintended use case of \toolname~can only be detected when it has been deployed within a real-world software development project. However, we argue that a low-code tool such as ours should embrace flexibility and support high configurability with DSLs, rapid prototyping, and contextual parameter simulation so that the advanced software practitioners have the freedom to experiment in a sandbox-like environment.

Another important recommendation from [P16], as given below, was to clearly communicate and promote the general applicability of the tools to developers, rather than suggesting an exclusive focus on older adults. Adopting this broader positioning could help alleviate concerns regarding the tool’s relevance across diverse user groups. Emphasising the versatility of \toolname~in addressing a wide range of accessibility and adaptation needs may enhance its perceived value and encourage adoption in more varied development contexts.

\begin{quote}
    \textit{"Because we don't directly go for something like that [a tool specifically for developing s/w for seniors rather than generic users]. We would rather try to find something that has some community support and all that stuff. So I'm really talking about the marketing [for the proposed tool: \toolname]."}
\end{quote}

\subsection{Dev Expectation 05: Documentation Support}

\subsubsection{\textbf{Context:}}

In \toolname, we aimed to provide developers with a mechanism to document their adaptation rules and individual adaptation operations. We also intended for this documentation to be embedded into the generated source code as inline comments, thereby improving code readability by clarifying what changes were made and the rationale behind each adaptation.

\subsubsection{\textbf{Reality}}

Several participants appreciated the inclusion of comment definitions in \textit{Adapt DSL} and the subsequent injection of these comments into the source code ($n=5$). Moreover, participants [P3] and [P7] viewed this feature as a valuable opportunity to upskill developers in accessibility standards, such as WCAG \cite{wcag2.2} and ISO 9241-171 \cite{iso9241}. By requiring developers to articulate the purpose of adaptations, the tool encourages critical reasoning, which may also benefit code review processes. [P7] made a particularly insightful observation, highlighting that the explicit reference to accessibility standards and the structured documentation process could enhance awareness of user experience considerations, especially among junior developers:

\begin{quote}
    \textit{"I find [that] there are lots of junior developers that don't go through web content accessibility documentation or guidelines or anything, and they just develop things, and often the user experience is part of just one UX developer's responsibility. Most of the time, developers neglect these requirements. So if we can have a tool like this, I think this will help even developers to understand certain requirements, certain user groups, and why they have do it."}
\end{quote}

\subsubsection{\textbf{Potential Solutions:}}

Aside from the positive aspects of our current documentation methods concerning \textit{Adapt DSL} discussed in the previous section, there were some minor criticisms ($n=4$) regarding the lack of documentation support in \textit{context DSL}. Participants noted the need for defining metadata that can describe various context nodes. This is particularly applicable when developers are allowed to introduce custom nodes. Such metadata would help ease the learning curve for new developers. Beyond explanatory documentation, [P3] pointed out an additional use case for metadata: defining end-user-facing error messages for specific contextual scenarios. For instance, if a user's mobile device has a malfunctioning microphone, a predefined warning embedded within the user’s context-of-use model could inform them that voice input functionality is unavailable in the generated adaptation due to hardware limitations.

In addition to developer-defined metadata, we, as low-code tool developers, can enhance documentation by providing supplementary materials. One feature requested by six participants was the ability to generate analytical summaries from the information modelled in \textit{context DSL}. In future iterations, the \textit{context DSL} editor could support a context parameter overview, coupled with intelligent suggestions. These suggestions could be based on predefined documentation and DSL templates aimed at addressing the most prominent accessibility issues identified through such analysis. For example, [P3] explained the following after suggesting the need for a bubble chart representation of contextual parameters in our context DSL run-time models.



\begin{quote}
    \textit{"when we're developing in general, if we know what type of requirements they have, and it's a heavy hitter: a thing that, let's say 50\% people have, [or] 80\% people have. [...] We probably want to do something more. And it's easier to... It's good to know what is there. It's also easier to remind ourselves [of user requirements]."}
\end{quote}


\subsection{Dev Expectation 06: Development Convenience through Code Generation}

\subsubsection{\textbf{Context:}}

Another recurring theme in the data concerned the enhanced convenience and user experience developers expect when integrating a new tool into their workflows. Our attempt to address this need through code generation received mixed feedback. Eleven participants responded positively to how aspects of \toolname~could improve their development convenience and overall experience, while a minority ($n=7$) offered critical feedback on the current code generation approach. Nonetheless, these critiques yielded valuable insights for potential future improvements.

\subsubsection{\textbf{Reality:}}

The most prominent theme in the positive feedback ($n=11$) was the perception that the code generation capabilities of \toolname~save developers time and effort. Participants noted that they would otherwise need to manually implement styling changes or complex source code modifications, such as integrating text-to-speech or speech-to-text modalities to accommodate individual users or varying contextual use cases. For example, participant [P6] articulated this sentiment clearly:

\begin{quote}
    \textit{"It's good, because otherwise, I have to write it [code] on my own. If something gets generated, it's always good."}
\end{quote}

Analysis of the negative feedback revealed two main concerns. The first relates to the increased cognitive load imposed on developers during the app development process. Anticipating and designing for diverse user needs, such as those of seniors, requires additional consideration of how the presentation and behaviour of the application should adapt across varying use cases. This is in addition to the already significant effort required to develop a static application. While this concern was raised by a minority of participants ($n=4$), it highlights a critical limitation in the current iteration of our code generation process.

The second concern pertains to code readability. Raised by three participants ($n=3$), this issue stems from the dynamic nature of our code generation architecture. Although we aim to preserve the original source code and derive variant versions to address age-related needs, the resulting complexity can impair developers' ability to read and understand their own code. This may be particularly problematic for onboarding junior developers, who could struggle to familiarise themselves with an evolving codebase. Participant [P8] illustrated this point effectively:

\begin{quote}
    \textit{"When you have many adaptations and then you actually lose the readability or explainability of your own source code."}
\end{quote}

\subsubsection{\textbf{Potential Solutions:}}

While these two issues could potentially offset the convenience gains offered by code generation, they are not insurmountable. To reduce the cognitive burden on developers, the low-code tool could be enhanced to provide context-aware adaptation suggestions based on already mapped user or user group parameters. For example, if the user context parameter \textit{vision impairment} is assigned the value \textit{macular degeneration}, the tool could recommend increasing visual contrast and offer links to relevant documentation or templates demonstrating how to implement such adaptations. Notably, seven participants explicitly requested a feature of this nature when asked for improvement suggestions. For instance, participant [P7] speculated on how the \toolname~could assist developers in better understanding user needs, particularly when end-users themselves may not fully recognise the extent of their accessibility challenges:

\begin{quote}
    \textit{"Sometimes, if we ask a user, do you have any impairment or disability?, they might come up with the one that comes to their head first. For example, they might say, Okay, I can't see red colour. I'm colour blind. So if we only use that, maybe they are missing something else. So, especially with this current generative AI models, if you can have some suggestions like, Okay, if this person has a red color impairment, maybe this person can have something else. I don't know. I'm just making things up, maybe this person [needs] help [with] yellow colour blindness too."}
\end{quote}

Another opportunity is to introduce an \textit{adaptation advisor} service that leverages historical adaptation rules, user context-of-use parameters, and application source code to automatically suggest adaptation rule models or partially completed DSL model templates. These recommendations would aim to address potential accessibility and personalisation barriers that the modelled user base may encounter. However, such predictions are inherently dependent/trained on historical DSL models and may introduce biases when the development context deviates from the one in which the advisor was trained. For example, AI-driven DSL model recommendations derived from use cases in the e-banking domain with a broad user base may not be appropriate for an e-health application targeting health professionals and cognitively impaired patients.


To address the second issue concerning code readability, we can introduce segmented code injection by inserting descriptive inline comments that clarify the changes made and the purpose of each adaptation. This approach was appreciated by several participants ($n=5$), yet the feedback suggests that further measures are needed to more distinctly separate generated code from developer-authored code.
By maintaining these adapted components in separate source files, they can be managed as independent entities within the project file structure. This decoupling also facilitates the generation of distinct file trees, each representing a different application permutation tailored to the accessibility and adaptation needs of specific user groups.

\subsection{Dev Expectation 07: Ease of Code Maintainability and Versioning}

\subsubsection{\textbf{Context:}}

Beyond the development phase, application maintenance and source code versioning represent critical stages in the software lifecycle. In this context, participants raised concerns about the challenges introduced by code generation, particularly in relation to maintaining the codebase alongside existing codebase maintenance and versioning practices. Accordingly, this section explores the specific challenges developers may encounter when using a low-code tool such as \toolname~and outlines the strategies they expect the tool to implement in order to mitigate these challenges effectively.

\subsubsection{\textbf{Reality:}}

Feedback from participants ($n=6$) revealed two key challenges associated with the proposed MDE-based approach. The first challenge concerns scalability. As various adaptive source code versions are generated to meet the diverse needs of end-users, the overall size and complexity of the codebase increase significantly. This leads to substantial maintenance overhead. Participant [P17] illustrated this concern with the following comment:

\begin{quote}
    \textit{"A certain page like that you have displayed here, we can implement it around 500 lines of codes. But in this kind of case [adaptation rule based code modifications], like sometimes, you have to go for 10,000 15,000 based on the different different scenarios, that is really hard, because that kind of code, basically is really hard to maintain."}
\end{quote}

The second issue relates to the practical management of the evolving codebase. As noted by [P1], an application may be maintained by hundreds of developers over its lifespan. In such cases, developers may independently define adaptation rules and generate varying application versions. Without a central governing mechanism, this could result in significant confusion and complexity in managing the codebase. An additional concern, raised by [P1] and [P10], involves the challenges of storing and versioning different source code permutations in repositories such as GitHub or GitLab. In particular, they noted the difficulty of handling merge conflicts between the master branch, which stores the original non-adapted version, and branches containing adapted versions for specific user contexts.

\subsubsection{\textbf{Potential Solutions:}}


The first strategy, as suggested by participants [P10], [P15], and [P17], is to advocate for targeting user groups rather than individual users when designing adaptive applications. Developers can already utilise our current modelling tools to define a generalised set of accessibility and adaptation needs applicable to user groups such as seniors with low vision, seniors with mild cognitive impairment, or seniors with arthritis. By focusing on user group-level adaptations, the number of app permutations required can be significantly reduced, leading to developers having to only maintain a few versions of the app instead of hundreds/thousands.

The second strategy, informed by feedback from [P10] and [P12], involves rethinking the current code generation architecture. Rather than generating entirely separate source code files for each adapted version, the original application can be designed to include placeholder UI widgets or components. By default, these placeholders render the non-adapted UI elements. When adaptations are triggered, either at design-time by developers or at run-time by end-users, these components can be dynamically replaced with context-appropriate widgets. [P10] illustrated this approach effectively:

\begin{quote}
    \textit{"In the Flutter you provide the space. So in that provided space, you'll be listening to the back end again. And with the adaptation coming, [.....] you're replacing the widget with the new widget coming according to the adaptation from the back [end]"}
\end{quote}

These two strategies simplify the management of adaptive application repositories. By reducing the number of adapted app instances or branches that developers must maintain, the likelihood of merge conflicts is lowered. Additionally, the ability to conditionally generate only specific components, while keeping adapted code decoupled from the core application logic, reduces the volume of code that developers must manage. This leads to improved maintainability overall.

\subsection{Dev Expectation 08: Ease of App Deployment}

\subsubsection{\textbf{Context:}}

In our prototype, we did not consider the practicalities involved in deploying an application to a user's device. However, this emerged as a significant concern that could influence developer adoption of tools similar to \toolname, as noted by several participants ($n=5$). For web applications, deployment tends to be straightforward, since most of the processing is handled server-side. In contrast, deploying mobile applications requires a more nuanced approach due to two main factors.

\subsubsection{\textbf{Reality:}}

In a real-world setting, not all end-users possess the latest mobile devices. This is particularly relevant for seniors, who often use older devices with limited processing power and storage. This issue was raised by [P17], whose team actively optimises application bundle size through strategies such as minifying third-party libraries. [P17] expressed concern that local storage of various generated adaptive code segments could result in a significantly larger codebase compared to that of a static, non-adaptive application.

The second concern relates to the nature of the deployment process on mobile platforms such as Apple's App Store and Google's Play Store. Participants [P12], [P14], [P15], and [P17] highlighted the lengthy and sometimes unpredictable approval processes required for application updates. Even if a centralised server is used to handle most code generation and distribute adapted app instances to users, the deployment bottleneck caused by app store policies remains. As [P15] noted:

\begin{quote}
    \textit{"The question of it being in an App Store is kind of a problem, right? Because you have to go through a lot of like getting approvals for app changes."}
\end{quote}

\subsubsection{\textbf{Potential Solutions}}

To address the limitations of mobile device resources, one solution is to offload code generation to a centralised server. When an adaptation rule, either defined by a developer or triggered by a user, is detected, the app can send an API call to a server-hosted code generator with the required parameters. The server can then generate the relevant code segments, return them to the device, and initiate a hot reload to apply the adaptations. The potential for API-based code generation was first suggested by participant [P10].


However, as previously discussed, platform store restrictions may introduce delays that undermine the timeliness of such updates. To mitigate this, adaptations can be delivered through \textit{Code Push} services, which support over-the-air (OTA) updates without requiring repeated app store approvals. These services enable developers to push source code updates directly to users’ devices. Nevertheless, they are subject to platform-specific policy constraints that limit the extent of permissible changes. Consequently, we anticipate the need to either integrate a commercial solution such as Shorebird into our model-driven engineering workflow or develop a custom code-push service tailored to our adaptation requirements. Participant [P12] highlighted this possibility by referencing Shorebird:

\begin{quote}
    \textit{"The only way you could potentially do it and have the code somewhere else is..., I don't know if you've seen `shore bird', it does over the air updates. So you release your app to the store, and then you can do patches, and you don't have to go through the store review process to get the patch. So you could potentially have a modified version of Shorebird that allows you to pull particular code for those data adaptations."}
\end{quote}


\subsection{Dev Expectation 09: Ease of Quality Assurance Tasks}

\subsubsection{\textbf{Context:}}

In our interview study, quality assurance (QA) was mentioned by only a minority of participants ($n=4$). Nevertheless, QA remains a critical component of the software development lifecycle and is therefore worth examining to understand the potential challenges that a tool like \toolname~may encounter in practice.

\subsubsection{\textbf{Reality:}}

The first issue, raised by [P1], [P12], [P16], and [P18], concerns the increased complexity of testing due to the presence of multiple adapted instances of the application. Testing each potential scenario to ensure reliable behaviour across all adaptations would be both time-consuming and resource-intensive. The second issue, identified by [P16] and [P18], relates to test automation. Since the user interface dynamically adapts to different user contexts, it does not remain consistent across scenarios. As a result, writing automated test scripts for complex UI adaptations, such as the integration of text-to-speech or speech-to-text functionality and changes in form navigation, becomes challenging for QA engineers. To further evidence this, let us consider the following statement from [P16]:

\begin{quote}
    \textit{"So one thing I'm a little worried about is that the testing complexity will obviously go quite high, so like in some of the cases, we mostly try to automate the testing. So, when you have several combinations to test, and that is the only thing that I'm a little bit worried about using [the demonstrated tool]."}
\end{quote}

\subsubsection{\textbf{Potential Solutions}}

Although participants did not offer concrete strategies for addressing the QA challenges, both [P1] and [P18] acknowledged that creating distinct test scenarios would be inevitable when generating personalised application instances. We propose that the resource-intensive nature of these testing activities can be mitigated by ensuring that the tool is designed not to interfere with existing automated testing workflows. For example, most presentation-level adaptations, which modify only the attributes of UI components, should not affect the underlying business logic or disrupt automated test scripts. However, more complex adaptations involving modality or workflow changes will likely require additional QA effort.

An encouraging and somewhat unexpected insight emerged from participants [P1] and [P13], who proposed that \toolname~could also function as a UI testing utility or sandbox during the development phase. For example, participant [P13] described a scenario in which developers could define adaptation rules to simulate how an app behaves under various platform parameter changes:

\begin{quote}
    \textit{"So having a set of [adaptation] rules defined initially and then even during the development, we can check how the different UIs are going to be when we change the rules, and how the UIs will fit into the screen and like the buttons and the text, because sometimes when you increase the text, they overflow from sides, sometimes out of the button and all that. So basically, like having a set of rules, and then we can quickly change which rule, and then we can see, like, how the UI is going to show in the screen, and then change to the second rule, see how the UI is going to look."}
\end{quote}

These simulation capabilities could help developers identify the most suitable UI presentation across varying conditions such as screen sizes, orientations, operating systems, and display features like True Tone or Night Shift on iOS \cite{iosDevGuidelines}. This positions \toolname~as a flexible and general-purpose UI testing tool, extending beyond age-specific accessibility needs.



\section{Discussion}\label{discussion}


\subsection{Summary of Potential Solutions}

{
    \renewcommand{\arraystretch}{1.3}
    \footnotesize
    
    \begin{longtable}{|>{\raggedright\arraybackslash}p{0.4cm}|
                      >{\raggedright\arraybackslash}p{2cm}|
                      >{\raggedright\arraybackslash}p{3.5cm}|
                      >{\raggedright\arraybackslash}p{6.5cm}|}
    
        \caption{A summary of potential general-use solutions the low-code developers could take to make their MDE tools more accessible to developers} \label{tab:potential_solutions} \\
        \hline
        \textbf{ID} & \textbf{Developer Expectation} & \textbf{Explanation} & \textbf{Potential Solutions} \\
        \hline \hline
        \endfirsthead
        
        \hline
        \textbf{ID} & \textbf{Developer Expectation} & \textbf{Explanation} & \textbf{Potential Solution} \\
        \hline \hline
        \endhead
        
        \hline
        \endfoot
        
        \hline
        \endlastfoot

            DE1 & The need to fulfil end-user accessibility and personalisation requirements & A majority of participants showcased a personal empathy towards the needs of seniors, indicating that they have a user-centric mindset and they are willing to adopt tools that let them build better apps. & 
                \begin{itemize}[left=0pt, label=--, nosep]
                \item A low-code tool that supplements current development workflows to generate adaptive and accessible apps using DSLs and MDE processes similar to \toolname.
                \item Should be accessible to developers as an open-source or commercial product.
                \end{itemize} \\ \hline
            
            DE2 & Easy Collection and Use of Accessibility Data & Any contextual data necessary to model the accessibility needs of end-users should not come at the cost of a negative user experience or user attrition. Additionally, modelling context data within the low-code tool should not be repetitive or time-consuming. & 
                \begin{itemize}[left=0pt, label=--, nosep]
                \item Minimal onboarding data collection (e.g., age, impairments).
                \item Supporting voluntary accessibility data input by users via a context information dashboard after onboarding
                \item Clear communication of data usage to address user privacy concerns.
                \item A universal accessibility profile defined by users and applied across applications (e.g., government services, or closed app ecosystems such as Apple or Amazon).
                \item Use of structured formats (e.g., XML/JSON) to support easy import into user context-of-use models
                \item Enable developers to model either user groups or individual users, allowing them to balance personalisation depth with development effort based on project requirements.
                \end{itemize} \\ \hline
            
            DE3 & Ease of use in modelling tools & Developers appreciate how easy the modelling tools are to use, as their intuitiveness and familiarity reduce adoption barriers such as steep learning curves. & 
                \begin{itemize}[left=0pt, label=--, nosep]
                \item Use familiar structures, such as tree hierarchies and JSON-style objects, within DSLs to ensure intuitive model design and navigation.
                \item Incorporate minor graphical enhancements, such as intuitive, self-explanatory icons, into semi-graphical DSLs such as context DSL to improve model clarity and readability.
                \item Provide dual support for textual/semi-graphical and graphical representations of DSLs to accommodate both technical and non-technical team members. Ensuring synchronisation between the two formats (e.g., through Xtext/Sirius integration \cite{sirius+xtext2017}) can facilitate smoother inter-role collaboration.
                \end{itemize} \\ \hline
           
            DE4 & Utility of MDE tools & Developers are more inclined to adopt a tool that is both flexible in its uses and has more utility than a tool that can only cater to the needs of one user group, thereby limiting its use cases (e.g., seniors) & 
                \begin{itemize}[left=0pt, label=--, nosep]
                \item Design the tool to support high configurability through flexible DSL definitions, allowing advanced developers to customise app behaviour more freely.
                \item Support a sandbox-like environment within the low-code platform where developers can simulate contextual parameters (e.g., user, platform, and environmental variables) and adaptation rules to rapidly prototype and preview app variations.
                \item Clearly communicate the broader applicability of the tool to address the needs of various user groups to avoid the perception that it is limited to a specific end-user demographic (e.g., older adults).
                \end{itemize} \\ \hline

            DE5 & Documentation support & Developers appreciate the ability to add metadata into their software artefacts, such as DSL models and source code. This practice encourages critical reasoning, improves artefact readability, and may aid in activities such as onboarding and code reviews. & 
                \begin{itemize}[left=0pt, label=--, nosep]
                \item Support developer-defined metadata within DSLs. Enable developers to document the purpose and intended outcomes of adaptation logic, which can assist in onboarding new team members and navigating complex low-code workflows.
                \item Automatically inject developer-defined metadata into generated source code as comment segments to enhance readability and traceability of adaptation logic.
                \item Allow practitioners to use DSL model metadata to define user-facing messages, such as contextual warnings or accessibility limitations due to device or environmental constraints (e.g., privacy prompts, notifications about missing hardware features).
                \item Provide a built-in feature to generate analytical summaries of modelled parameters (e.g., bubble charts visualising the most prominent impairments) to help developers prioritise user needs more effectively.
                \end{itemize} \\ \hline

            DE6 & Development convenience through code generation & Code generation should make the application development more convenient to developers by automating repetitive or resource-intensive tasks. However, low-code tool developers should ensure that the generated code does not compromise the readability, structure, or maintainability of the overall codebase, as doing so may negate the benefits of automation. & 
                \begin{itemize}[left=0pt, label=--, nosep]
                \item Provide context-aware adaptation suggestions based on mapped user or user group parameters, guiding developers with intelligent and actionable recommendations and relevant DSL template links.
                \item Clearly separate tool-generated code from developer-authored code to prevent confusion and improve traceability
                \item Generate separate project file trees for each user group or application permutation, making it easier to manage, test, and distribute personalised versions.
                \end{itemize} \\ \hline

            DE7 & Ease of code maintenance and versioning & Developers expect low-code tools such as \toolname~to support scalable code management and integrate seamlessly with existing version control systems such as GitHub and GitLab. As adaptive code variants increase, the tool should help reduce maintenance overhead and avoid codebase complexity. Developers also highlighted the risk of merge conflicts when multiple permutations of an app are maintained across different branches and contributors. & 
                \begin{itemize}[left=0pt, label=--, nosep]
                \item Encourage user group-level adaptations (e.g., seniors with low vision or arthritis) instead of individualised versions to minimise the number of app permutations for projects with a large user base.
                \item Use placeholder UI components in the base application that can be dynamically replaced with context-appropriate widgets when adaptations are triggered.
                \item Avoid generating entirely separate source files for each adaptation; instead, support modular replacement of components to keep adapted code decoupled from core logic.
                \item Reduce the number of parallel app branches to lower the likelihood of merge conflicts and simplify repository management.
                \end{itemize} \\ \hline

            DE8 & Ease of app deployment & Deploying mobile applications present unique challenges compared to web or desktop environments. These include limited device resources, particularly on older hardware, and restrictive app store policies that delay the delivery of dynamic updates. Developers expect a low-code tool such as \toolname~to provide seamless deployment mechanisms that address these limitations while remaining compliant with app store constraints. & 
                \begin{itemize}[left=0pt, label=--, nosep]
                \item Use API-based code generation offloaded to a centralised server, allowing adaptations to be generated remotely and delivered to the app, and be applied to the app at run-time via a hot reload. This reduces app bundle size and minimises storage usage on user devices.
                \item Integrate support for commercial Code Push services (e.g., Shorebird) to enable over-the-air (OTA) delivery of updates, avoiding repeated app store review cycles.
                \item If necessary, develop a custom, app store policy-compliant code push mechanism tailored to the specific needs of low-code app adaptation workflows.
                \end{itemize} \\ \hline

            DE9 & Ease of quality assurance & A low-code approach that generates adapted app instances will inevitably result in numerous test scenarios, especially when aiming for high levels of personalisation. Additionally, dynamic adaptations may complicate the creation of automated test scripts. Developers, therefore, expect the low-code tool to assist in quality assurance rather than introducing additional complexity. & 
                \begin{itemize}[left=0pt, label=--, nosep]
                \item Design the tool to be compatible with existing automated testing workflows, ensuring that simple presentation-level adaptations (e.g., UI attribute changes) do not interfere with business logic or disrupt automated test scripts.
                \item Acknowledge that complex adaptations involving modality or workflow changes may require additional testing but should be isolated to minimise their impact on the broader testing process.
                \end{itemize} \\ \hline

    \end{longtable}
}

\subsection{Implications for Practice}

\subsubsection{\textbf{Practical Viability of Low-Code tools in Developing Accessible and Adaptive Apps}}

Our findings suggest that a low-code tool leveraging domain-specific language (DSL) models to capture the accessibility and personalisation needs of senior end-users and subsequently transform these models into executable application instances through code generation is a promising and viable approach. This conclusion is supported by the positive reception of our prototype, with 17 out of 18 participants indicating their willingness to adopt such a tool, provided it meets the developers' expectations outlined in Section~\ref{results}. As further evidence, participant [P18] remarked favourably on the demonstrated prototype and noted that a tool of this nature could be a natural solution in real-world development settings when accessibility needs emerge:

\begin{quote}
    \textit{"I think, for a proof of concept, this is very good. And I can even see, like, if this is something that arose from the internal need, I can imagine there are companies which use a similar approach to this, because they came up with it, because they needed a specific solution. This seems very like down-to-earth, practical solution. I just think if you want to deploy, if you want to convince others to use it, it just needs more polish. But as a concept, I think it's a nice concept. It's good."}
\end{quote}

However, our investigation into the state of the art indicates that UI adaptations through model-driven engineering remain an underexplored area within low-code research \cite{wickramathilaka2023, wickramathilaka2025technical}. Existing implementations are predominantly limited to proof-of-concept prototypes, including our own tool, \toolname. Nevertheless, our work consolidates the strengths of prior approaches while introducing novel contributions \cite{wickramathilaka2025technical}. This creates a strong foundation upon which tool developers can build to produce a more polished, open-source or commercial solution designed for practical use by real-world software development teams.

\subsubsection{\textbf{Implications for Future Low-Code Tool Developers}}

In extending an invitation to low-code tool developers to address this gap in the software industry, we strongly encourage the incorporation of the lessons learned from our interview study with Flutter developers, as discussed in Section~\ref{results}. Furthermore, we synthesise these insights into a set of general-purpose tooling recommendations presented in Table~\ref{tab:potential_solutions}, which may serve as practical guidelines during the development process. Adhering to these recommendations can increase the likelihood of producing a tool that aligns with the expectations and needs of real-world software practitioners, thereby reducing adoption barriers. We believe that following these guidelines could lead to the creation of a practical low-code tool capable of dynamically generating accessible and adaptive application instances at scale, while also supporting these variations across the software development lifecycle: from requirement elicitation through to design, development, deployment, and ongoing maintenance.

\subsubsection{\textbf{Implications for Developing More Accessible and Personalised Apps}}

In Table~\ref{tab:potential_solutions}, the first developer expectation highlights the motivation among software practitioners to build more accessible and adaptive applications that address the needs of their end-users. While this motivation encompasses both personal and professional perspectives, it can also be viewed through a pragmatic, future-focused lens. By designing and adopting tools such as \toolname~to eliminate accessibility barriers and support personalisation, we as software practitioners are not only addressing current user needs but also anticipating our own future requirements (or responding to age-related challenges we may already be experiencing). Given the growing dependence on a wide range of applications across multiple platforms, this approach represents a proactive investment in creating inclusive digital environments for all, including our future selves. Perhaps an introspection about the future from a young but experienced app developer ([P1]) best illustrates this viewpoint:

\begin{quote}
    \textit{"This might be a must-use model or approach in the future, when our current generation becomes older, with certain issues in our real life, and then we'll still be using apps to order stuff or whatever or to connect with our friends. We will need to look... we will need to have different capabilities [to address age-related needs] in our applications."}
\end{quote}


\subsection{Limitations}

\subsubsection{\textbf{Lack of Non-Developer Participants}}

Our interview study included only software engineers as participants. However, software development teams typically involve a range of non-developer practitioners, such as project managers, requirements engineers, quality engineers, and user experience specialists. Including these roles may have revealed additional insights into the enablers and barriers faced when adopting a low-code approach for developing accessible and adaptive applications.

\subsubsection{\textbf{Flutter Developer Recruitment}}

Recruiting participants with at least six months of professional Flutter experience was challenging, likely due to the framework's relatively recent 1.0 release in late 2018~\cite{flutter}. Nevertheless, we successfully recruited 16 developers with varying levels of Flutter and industry experience. Additionally, two more participants were included despite not having professional Flutter experience, both due to their familiarity with the framework and substantial expertise in designing software for older adults. These participants were also well-versed in accessibility standards such as WCAG~\cite{wcag2.2} and ATAG~\cite{atag}.

\subsubsection{\textbf{Tool Development Environment}}

Currently, our \toolname~tool prototype has been developed within an Eclipse IDE-based development environment, as most of the dependencies critical to our DSL and MDE component development are only available on the Eclipse Modelling Framework (EMF) \cite{emf}, such as Xtext \cite{xtend}, Xtend \cite{xtend}, Sirius \cite{sirius}, and Acceleo \cite{acceleo}.  

\subsubsection{\textbf{Tool Demonstrations via Video Presentation}}

Although the current \toolname~prototype is functional and could be installed on participant devices for direct use, the installation process remains complex. It requires the installation of the Eclipse IDE, multiple EMF-related plugin dependencies, various configurations, and a fresh Flutter setup within Eclipse. These steps would be overly time-consuming for participants, particularly as they are working professionals volunteering for the study out of interest. Therefore, we opted for a video demonstration approach to effectively communicate the core functionality of the \toolname~toolsuite. Nonetheless, it is plausible that hands-on experimentation would have revealed more nuanced feedback regarding tool usability and improvement opportunities.

\subsection{Future Work}

\subsubsection{\textbf{Iterative Improvements to \toolname}}

We plan to iteratively implement the potential solutions identified to meet the expectations of our developer participants. The next iteration will prioritise enhancements in code generation, maintainability, and reusability. Following that, we will focus on adaptive app deployment and practical methods for collecting user context data. The third iteration will address improvements in quality assurance. If development reaches this stage, we aim to release \toolname~as an open-source plugin built on the Eclipse development environment with the potential to be extended into a more ubiquitous environment, such as VSCode.

\subsubsection{\textbf{Packaging \toolname~for Easy Installation and Facilitating Hands-on Experimentation}}

Following iterative improvements, we plan to package \toolname~alongside its dependencies to ensure a more streamlined installation process. This will be accompanied by a detailed wiki outlining installation steps and usage guidelines. Subsequently, we aim to conduct another evaluation phase involving a group of developers, allowing them to engage with the tool over an extended period. This hands-on engagement is expected to yield richer insights into actual tool usage, surpassing those obtainable through video demonstrations alone.

\subsubsection{\textbf{Extended Evaluation With Diverse Software Practitioners}}

Following these improvements, we intend to conduct an extended user study involving a broader range of software practitioners, including UX engineers, software engineers, quality engineers, requirements engineers, domain experts, and other relevant stakeholders. This would help identify further strategies to reduce barriers to adopting \toolname~within software teams, ultimately promoting the development of more accessible and adaptive applications for disadvantaged user groups such as older adults.


\section{Related Work}



The selection criteria for including studies in this comparative analysis are as follows: (1) the paper must propose a model-driven engineering (MDE) approach; (2) it must consider accessibility and/or personalisation needs of end users; and (3) the target users of the proposed tool must be software practitioners. Based on these criteria, we identified 14 relevant studies and analysed them with respect to the following aspects: (1) the nature of the contribution (e.g., whether a concrete prototype or simplified proof-of-concept was presented); (2) the presence and nature of evaluations conducted; and (3) whether explicit recommendations were made for the low-code developer community.

A majority of the studies (8 out of 14) did not include user-study-based evaluations and therefore offer limited insights into the real-world applicability of the proposed tools for their intended users~\cite{rieger2020, dias2021, krainz2016, watanabe2010, ghaibi2017, bacha2011, braham2022, minon2015}. These studies either provided no evaluation or employed only proof-of-concept examples or case studies.
A second group (4 out of 14) conducted evaluations solely with end users ~\cite{bendaly2018, yigitbas2020, bongratz2012, akiki2016}: individuals who ultimately interact with applications developed using these tools. The findings from these evaluations allowed the authors to comment on the generalisability and suitability of the proposed tools for the intended end-user groups. The final and least common category (2 out of 14) comprises studies that evaluated the tools from the perspective of software practitioners, i.e., the primary users of the MDE tools~\cite{bouraoui2019, krainz2018}. We argue that this perspective is critical for providing actionable guidance to the low-code development community.

Unfortunately, our review did not identify any explicitly articulated, concrete recommendations or insights directed at fellow low-code developers. We posit that this absence is largely attributable to the underexploration of the software practitioner or modeller perspective within the model-driven engineering (MDE) literature concerning accessibility and personalisation.
Of the two studies that did engage with this perspective, the work by Bouraoui and Gharbi~\cite{bouraoui2019} primarily focused on validating their proposed approach, offering limited discussion on how the insights might inform the practices of other low-code developers pursuing similar objectives. Similarly, the study by Krainz et al.~\cite{krainz2018} represents a missed opportunity; their evaluation involving software developers was confined to a quantitative survey, yielding only limited insights into the practitioner experience.

Nevertheless, these studies contribute valuable approaches for integrating MDE principles with accessibility and personalisation concerns to improve software application experiences for diverse user populations. For low-code tool developers seeking to draw lessons from these works, it is necessary to engage deeply with the respective technical artefacts, such as DSL metamodels, editor implementations, prototype applications, and generated source code examples, to extract relevant design knowledge.
In this context, our objective is to support low-code developers by sharing the lessons learned throughout our project. Hence, in this paper, we propose a set of design considerations and recommendations aimed at facilitating the development of MDE tools tailored for software practitioners. These insights are aimed at assisting future low-code tool developers in avoiding the need to initiate their efforts from scratch, as was necessary in our case.

\section{Summary}\label{summary}


In this paper, we investigated the challenge of low-code support for more accessible apps for the elderly. We examined empirical insights derived from an interview study involving 18 software developers who evaluated a model-driven engineering (MDE) tool we developed, named \toolname. This tool captures the accessibility and personalisation requirements of senior users using two domain-specific languages (DSLs). It then combines these models with a non-adapted base application to generate multiple adapted application instances, requiring minimal intervention from developers.
We used our insights from this empirical study to derive some key findings. First, we identify several key expectations that developers had when using \toolname~as a supplementary tool across various phases of the software development lifecycle. Next, we contextualise these expectations using empirical data, examining the extent to which \toolname~meets them and identifying areas for improvement. Finally, we present a set of generalised design recommendations for low-code tool developers aiming to address the challenges of accessibility and adaptivity in software applications through MDE. These recommendations are grounded in our empirical findings and intended to support the development of industry-grade low-code tools informed by real-world software practitioner needs.

\section{Acknowledgments}

Authors are supported by   Australian   Research   Council (ARC) Laureate Fellowship FL190100035. Haggag is supported by a National Intelligence Post-doctoral Fellowship. Our sincere gratitude goes to the participants who took part in the user study.

\bibliographystyle{ACM-Reference-Format}
\bibliography{bibliography}  


\end{document}